\documentclass[aps,prd,twocolumn,superscriptaddress,showpacs,nofootinbib]{revtex4-2}

\usepackage{graphicx} % 用于插入图片
\usepackage{amsmath} % 用于数学公式
\usepackage{amsfonts} % 用于数学字体
\usepackage{amssymb} % 用于数学符号
\usepackage{url} % 用于处理URL
\usepackage{hyperref}
\usepackage{color}
\usepackage{array}

\begin{document}

% 标题
\title{Probing Circumstellar Material and Shock Acceleration in Core-Collapse Supernovae with High-Energy Neutrinos}

% 作者信息
\author{Yi-Long Duan}
\affiliation{School of Mathematics and Physics, China University of Geosciences, Wuhan 430074, China}

\author{Tuohuniyazi Tuniyazi}
\affiliation{School of Mathematics and Physics, China University of Geosciences, Wuhan 430074, China}

\author{Gang Guo}
\email{Corresponding author: guogang@cug.edu.cn} 
\affiliation{School of Mathematics and Physics, China University of Geosciences, Wuhan 430074, China}
\affiliation{Shenzhen Research Institute of China University of Geosciences, Shenzhen 518057, China}

% 摘要
\begin{abstract}
We study high-energy (HE) neutrino production from interactions between supernova (SN) ejecta and the surrounding circumstellar material (CSM), focusing on regular Type~II and Type~IIn SNe. Using observationally inferred CSM density distributions, we calculate the resulting neutrino fluxes and examine their dependence on key parameters, including the CSM density normalization $D_*$, outer radius $R_{\rm csm}$, proton acceleration efficiency $\epsilon_p$, and magnetic energy fraction $\epsilon_B$. Detection prospects are assessed with a binned likelihood analysis for IceCube, indicating that nearby SNe with moderately dense, confined CSM can produce detectable signals, with a typical detection horizon of $\sim 0.1$--1~Mpc. For a Galactic SN at $\sim 10$~kpc, high-statistics neutrino data with detailed temporal and spectral information can constrain $D_*$, $R_{\rm csm}$, and $\epsilon_p$ to within a factor of $\sim 10$ or to a precision of $\sim 20\%$, depending on the assumed values of $D_*$ and $R_{\rm csm}$. These neutrino signals thus provide a complementary probe of the CSM profile and shock acceleration, alongside traditional electromagnetic observations.
\end{abstract}

% 关键词
%\pacs{03.65.-w, 03.67.-a, 03.65.Ta} % 物理学分类代码

% 正文
\maketitle

\section{Introduction}

Core-collapse supernovae (CCSNe), among the most energetic explosions in the Universe, mark the violent deaths of massive stars and are copious sources of low-energy (LE) neutrinos in the range of 10--100 MeV, produced during core collapse and proto–neutron-star cooling. These neutrinos carry direct information about the explosion mechanism, neutrino interaction and transport, and the physics of hot and dense nuclear matter \cite{Janka:2006fh,Janka:2012wk,Burrows:2013,Janka:2017vlw,Fischer:2023ebq,Suzuki:2024fse,Raffelt:2025wty}. In addition to this well-established LE neutrino signal, CCSNe have been suggested as potential sources of HE neutrinos in the TeV–PeV range \cite{Tamborra:2018,Tamborra:2024fcd}. Several HE neutrino production mechanisms have been explored. For instance, mildly relativistic jets may be launched during core collapse and could provide sites for proton acceleration at shocks as well as HE neutrino production \cite{Meszaros2001,Razzaque_2003,Razzaque2004,Ando:2005xi,RAZZAQUE_2005,Horiuchi:2007xi,Bartos:2012sg,Murase2013,Fraija2013,Xiao2014,Bhattacharya2014,Varela2015,Tamborra2015,Senno2015,Senno:2017vtd,Denton:2017jwk,Denton:2018tdj,He:2018lwb,Guo:2022zyl,Chang:2022hqj,Guarini:2022hry}. An alternative scenario involves rapidly rotating compact remnants, the nascent pulsars or magnetars, which can drive relativistic winds that accelerate charged particles and generate HE neutrinos via inelastic $pp$ scatterings when the accelerated protons collide with the expanding SN ejecta \cite{Zhang:2002xv,Nagataki:2003kq,Murase_2009,Fang:2014qva}.

Another scenario involves interactions between the SN ejecta and the CSM surrounding the progenitor. As the ejecta collide with the CSM, strong collisionless shocks develop and can accelerate protons to high energies. Inelastic $pp$ collisions between the accelerated protons and the CSM then produce charged pions and kaons, whose decays yield HE neutrinos \cite{Murase:2011}. This ejecta–CSM interaction model has attracted increasing attention in recent years, as a substantial fraction of CCSNe are inferred to undergo enhanced or eruptive mass loss shortly before explosion, leading to the formation of dense CSM around their progenitors, as revealed by multiwavelength observations of young SNe \cite{Gal-Yam:2014tua,Yaron:2017}.

The prospect of detecting HE neutrinos from interacting SNe has been addressed in numerous studies. Early studies demonstrated that Type IIn SNe with extremely dense CSM could produce multi-TeV neutrinos detectable within $\sim$20–30 Mpc by current neutrino telescopes \cite{Murase:2011}. Subsequent studies modeled the time-dependent SN–CSM interaction and suggested that Type IIn SNe could contribute significantly to the diffuse HE neutrino background \cite{Zirakashvili:2015mua,Petropoulou:2017ymv}. The interaction scenario was later extended to other SN types, including Type IIP, IIL, and Ibc, which have less dense but still substantial CSM, implying that the neutrino signals from SNe in the Milky Way and nearby galaxies could be observable \cite{Murase:2017pfe, Kheirandish:2022eox}. More recent efforts have focused on specific events, such as the nearby Type II SN 2023ixf \cite{Kheirandish:2022eox,Guetta:2023mls,Kimura:2024lvt,Cosentino:2025sdd,IceCube:2025lzi}, as well as potential associations between IceCube neutrino events and interacting SNe, including SN 2023uqf \cite{Stein:2025uxi} and the Type IIn SNe 2023syz and 2025cbj \cite{Lu:2025jks,Garrappa:2025ves}.

Most existing studies of HE neutrino emission from the SN ejecta–CSM interactions adopt specific or benchmark choices for the CSM profile and shock properties, whereas a systematic exploration of a wide range of physically plausible parameters remains limited. In particular, the dependence of neutrino signals on key physical ingredients--such as the CSM density normalization, radial extent and slope, and the microphysics of collisionless shocks including particle acceleration and magnetic-field amplification--has not been thoroughly investigated. More importantly, the inverse problem, i.e., whether neutrino observations themselves could provide meaningful constraints on the CSM structure or shock physics, remains largely unexplored.

Motivated by these considerations, we study HE neutrino emission from nearby CCSNe, explicitly accounting for the physically allowed ranges of key parameters. This approach captures the diversity of SN populations and allows us to assess how variations in CSM properties and shock microphysics shape the resulting neutrino spectra, light curves, and event rates. Although the Galactic CCSN rate is modest ($\sim$ 1--3 per century \cite{Adams:2013ana,Rozwadowska:2020nab}), a nearby event could produce tens to hundreds of TeV–PeV neutrino events, enabling detailed temporal and spectral analyses. For a future Galactic Type II SN, we explore the potential of HE neutrino signals as complementary probes of both the CSM density profile and the physics of collisionless shocks, incorporating parameter uncertainties.

The structure of this paper is as follows. In Sec.~\ref{sec:basic}, we introduce the profiles of the SN ejecta and the CSM, including  the distribution of mass-loss rates inferred from observations. We then review neutrino production in the SN ejecta–CSM interaction model, compute the resulting HE neutrino fluxes, and discuss their dependence on the relevant physical parameters. In Sec.~\ref{sec:detection}, we examine the detection prospects of these HE neutrinos, both from individual nearby SNe and from the cumulative diffuse flux. In the case of a Galactic SN with high event statistics, we explore in Sec.~\ref{sec:constrain} the potential of using neutrino observations to constrain the CSM density profile and shock acceleration. Finally, we summarize our conclusions in Sec.~\ref{sec:summary}.

\section{Basic formalisms}
\label{sec:basic}

\subsection{the profiles of SN ejecta and the CSM}

In this work, we focus on Type II SNe with hydrogen-rich envelope. We assume that the kinetic energy of the SN ejecta interacting with the CSM is $\mathcal{E}_\text{ej} = 10^{51} \text{ erg}$, and the ejecta mass is $M_\text{ej} = 10M_\odot \approx 1.989 \times 10^{34} \text{ g}$. Numerical simulations of SN explosions indicate that the SN ejecta is divided into an inner part ($\rho_\text{ej} \propto r^{-\delta}$) and an outer part ($\rho_\text{ej} \propto r^{-n}$) \cite{Matzner:1998mg}, where $\delta$ and $n$ are the slope parameters. 
The outer slope depends on the progenitor structure and typically lies in the range $n \approx 10$--12 for CCSNe, while the density of the inner core is nearly flat with $\delta \sim 0$--1 \cite{Moriya:2013hka}.

Up to a characteristic radius $R_{\rm csm}$, the CSM density can be described by
\begin{equation}
\rho_{\rm csm}(r) = D r^{-w},  
  \label{eq:rho}
\end{equation}
where $w$ is the density slope. Throughout this work, we focus on the wind-like case with $w=2$, while keeping the explicit dependence on $w$ in the general expressions below. In this case, the parameter $D$ is constant and directly related to the steady mass-loss history of the progenitor prior to core collapse,
\begin{align}
D ={\dot{M} \over 4\pi V_{\rm w}},
\end{align}
where $\dot{M}$ is the mass loss rate and $V_{\rm w}$ the wind velocity. Following Ref.~\cite{Murase:2017pfe}, we introduce the dimensionless parameter
\begin{align}
D_* & \equiv {D \over 5 \times 10^{16} \rm{~g~cm^{-1}}} \nonumber \\
& \approx {\dot M \over 0.1~M_\odot~{\rm yr}^{-1}} {100~{\rm km~s^{-1}} \over V_{\rm w}}, \label{eq:D-mass-loss}
\end{align}
to quantify the magnitude of the CSM density. For representative values
$\dot{M} = 0.1 M_\odot~{\rm yr}^{-1}$ and $V_{\rm w}=100~{\rm km~s^{-1}}$, we obtain $D_* \approx 1$.

The values of $D_*$ span several orders of magnitude across different SN subclasses, reflecting large diversity of mass-loss histories shortly prior to SN explosion.
Observationally, the properties of the CSM—including its density, velocity, as well as its radial extent—are inferred using a combination of diagnostics, such as early-time spectroscopy, persistent narrow emission lines, optical and ultraviolet excess emission, light-curve rise times and shapes, as well as radio and x-ray observations. Type~IIn SNe are characterized by persistent narrow hydrogen emission lines, which directly trace slow-moving, dense CSM photoionized or shock-heated by the SN explosion. These features, together with luminous and slowly evolving light curves, indicate sustained and strong interaction between the SN ejecta with an extended, dense CSM. Modeling of the optical spectra, bolometric light curves, and x-ray emission typically requires high progenitor mass-loss rates of $\dot{M} \sim 10^{-3}$--$1\,M_\odot\,\mathrm{yr^{-1}}$ for wind velocities $V_{\rm w} \sim 100$--$1000\,\mathrm{km\,s^{-1}}$, maintained over years to decades prior to core-collapse \cite{Smith:2014txa,Kiewe:2012,Taddia:2013nga,Moriya:2014,Sarmah:2022vra,Ransome:2024cza,Salmaso:2024jry}. This corresponds to typical CSM radii $R_{\rm CSM} \sim 10^{15}$--$10^{16}\,\mathrm{cm}$ or larger. The resulting $D_*$ lies in the range of 0.01--10 assuming $V_{\rm w}=100$~km~s$^{-1}$. In contrast, red supergiant (RSG) progenitors of regular Type~II SNe are generally inferred to have much lower steady mass-loss rates, $\dot{M} \sim 10^{-6}$--$10^{-5}\,M_\odot\,\mathrm{yr^{-1}}$, when averaged over long timescales \cite{Smith:2014txa}. However, an increasing number of early-time optical, ultraviolet, millimeter and flash-spectroscopy observations now consistently indicates that a significant fraction of regular Type~II SNe (mainly IIP and IIL) undergo short-lived episodes of strongly enhanced mass loss in the final months to $\sim$year before explosion (see, e.g., \cite{Khazov:2016,Yaron:2017,Morozova:2017,Morozova:2018,Forster:2018,Moriya:2018,Jacobson-Galan:2024,Bruch:2021,Bruch:2023,Jacobson-Galan:2024,Silva-Farfan:2024,Hinds:2025qia}). These eruptive episodes can reach mass-loss rates comparable to those inferred for Type~IIn SNe. However, because they are short-lived, the resulting dense CSM is confined to compact radii, $R_{\rm CSM} \sim 10^{14}$--$10^{15}\,\mathrm{cm}$. Detailed analyses of recent nearby events, such as SNe~2020fqv, 2023ixf, and 2024ggi, similarly point to very large values of $\dot{M}$ and compact CSM extents of order a few $\times 10^{14}\,\mathrm{cm}$ \cite{Tinyanont:2022,Shrestha:2024,Hu:2024wkh,Hu:2025}. For normal Type II SNe, we only focus on the confined dense CSM produced during these eruptive mass-loss phases, which are expected to yield significantly enhanced HE neutrino fluxes.

In many previous studies, a single representative value of $D_*$ was typically adopted for each SN subtype. For instance, canonical values such as $D_* \sim 1$ for Type IIn SNe have been used to model HE neutrino emission \cite{Murase:2017pfe,Kheirandish:2022eox,Sarmah:2022vra}. To better capture the intrinsic diversity in progenitor mass-loss histories across SN populations, we instead adopt a more realistic approach by incorporating observationally inferred distributions of $\dot{M}/V_{\rm w}$. The inclusion of these empirical distributions allows us to statistically evaluate the discovery potential of HE neutrinos from nearby SNe and to estimate the diffuse HE neutrino background from all similar populations.

\begin{figure}[htbp]
\centering
\includegraphics[width=\columnwidth]{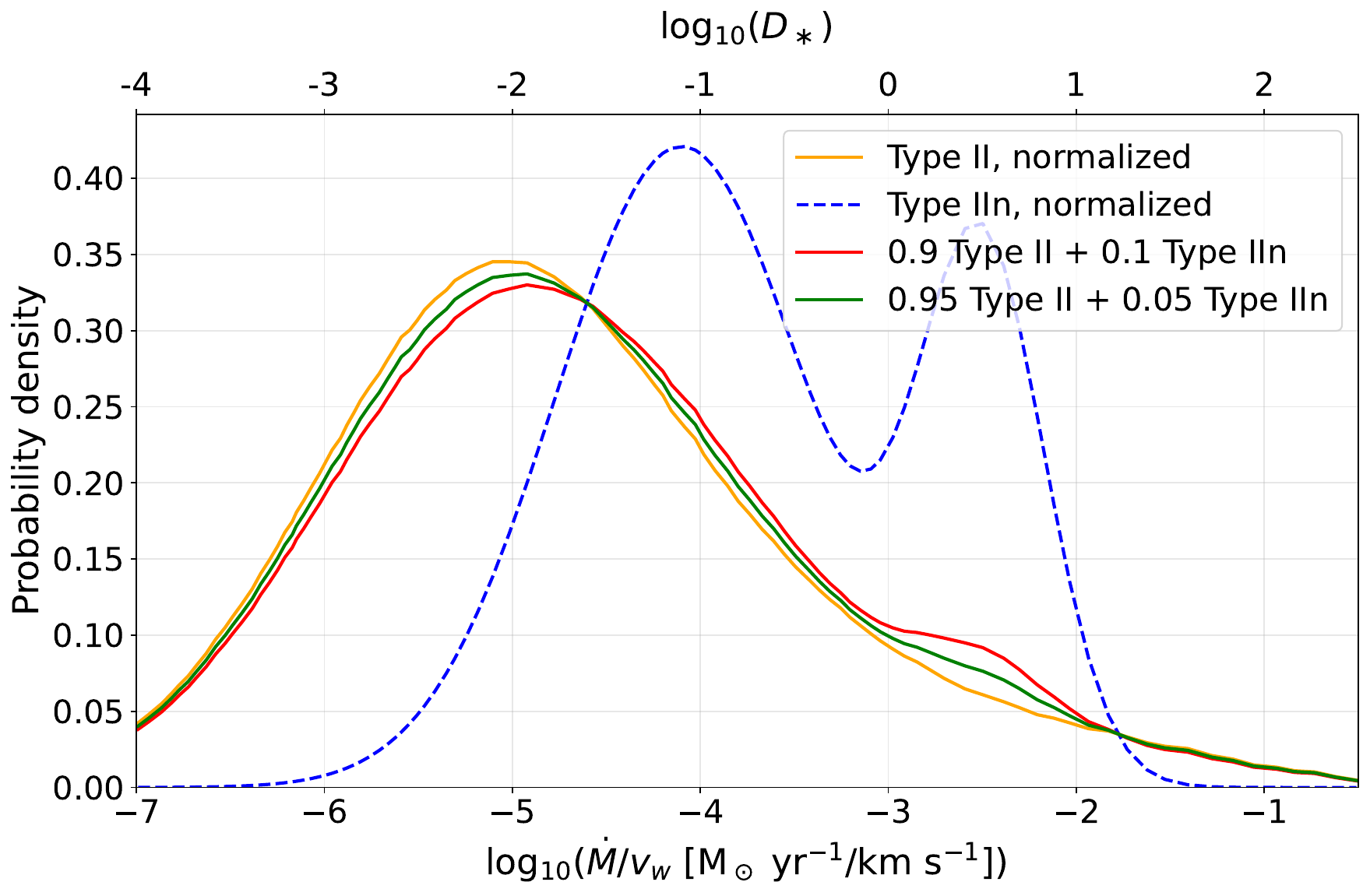}
\caption{Normalized (to unity) distributions of $\dot M/V_{\rm w}$, or equivalently, $D_*$ [see Eq.~\eqref{eq:D-mass-loss}], for regular Type II (solid orange) and Type IIn (dashed blue) SNe. Assuming relative fractions of 90\% (95\%) for Type II and 10\% (5\%) for Type IIn, the corresponding weighted distribution combining both subtypes is shown by the red (green) curve.}
\label{fig:distribution}
\end{figure}

Figure~\ref{fig:distribution} shows the distributions of $\dot{M}/V_{\rm w}$, or equivalently, $D_*$ [see Eq.~\eqref{eq:D-mass-loss}], adopted in this study for normal Type~II (excluding Type~IIn\footnote{Unless stated otherwise, the term “Type II” in this work refers exclusively to regular Type II SNe, excluding Type IIn events.}; left) and Type~IIn (right) SNe. For normal Type~II SNe, we use the distribution derived from early optical observations of a magnitude-limited sample of 639 Type~II SNe from the Zwicky Transient Facility~\cite{Hinds:2025qia} (see their Fig.~11 for the inferred $\dot{M}$, with $V_{\rm w}$ fixed at 10~km~s$^{-1}$). For Type~IIn SNe, we adopt a bimodal Gaussian distribution that fits the inferred values of $\dot{M}/V_{\rm w}$ obtained from systematic light-curve modeling of a sample of 57 Type~IIn SNe~\cite{Ransome:2024cza} (see their Fig.~15). Since Type~IIn SNe are much rarer than normal Type~II SNe, we also show weighted distributions that account for the relative fractions of the two subtypes (90\% or 95\% for normal Type~II and the remaining 10\% or 5\% for Type~IIn) for illustration. Notably, we find that the regular Type II SNe contribute predominantly even at the highest $\dot{M}/V_{\rm w}$ values.

In addition to $D_*$, the CSM outer radius $R_{\rm csm}$ plays a crucial role in shaping the HE neutrino signal. As discussed above, the CSM surrounding normal Type~II SNe is typically much more compact than that associated with Type~IIn SNe. Accordingly, we consider $R_{\rm csm}$ in the range $10^{14}$--$10^{15}\,\mathrm{cm}$ for normal Type~II SNe. For Type~IIn SNe, we consider more extended CSM configurations with $R_{\rm csm}$ spanning $10^{15}$--$10^{17}\,\mathrm{cm}$. Throughout this work, we adopt $R_\ast=6\times 10^{13}$~cm, $M_{\rm ej}=10 M_{\odot}$, $\mathcal{E}_{\rm ej}=10^{51}$~erg, $n=10$, and $\delta=0$, unless otherwise stated.

\subsection{neutrino production by ejecta--CSM interaction}

We closely follow the formalism of Ref.~\cite{Pitik:2023vcg} to compute neutrino fluxes from the SN ejecta–CSM interactions. For completeness, we briefly summarize below the key expressions relevant to neutrino production and refer the reader to Ref.~\cite{Pitik:2023vcg} for further details.

The interaction between the SN ejecta and the surrounding CSM produces a thin shocked shell, bounded by a forward shock propagating into the CSM and a reverse shock propagating into the ejecta~\cite{Chevalier:1982a,Chevalier:1982,Chevalier:1994,Moriya:2013hka}. We focus on particle acceleration and HE neutrino production at the forward shock, which is expected to dominate the emission \cite{Suzuki:2020qui}. 

The radius of the shocked shell evolves as \cite{Pitik:2023vcg}
\begin{align}
R_{\mathrm{s}}(t) = 
\begin{cases}
R_\star + \left[
\dfrac{(3-w)(4-w)g_n}{(n-4)(n-3)D} \right]^{\frac{1}{n-w}} t^{\frac{n-3}{n-w}}, & t \le t_{\mathrm{dec}}, \\
R_{\mathrm{dec}}\left( \dfrac{t}{t_{\mathrm{dec}}}\right)^{\lambda}, & t > t_{\mathrm{dec}},
\end{cases}
\label{eq:Rs}
\end{align}
where
\begin{align}
g_n = {1 \over 4\pi n} {[10(n-5) \mathcal{E}_{\rm ej}]^{(n-3)/2}
\over 
[3(n-3)M_{\rm ej}]^{(n-5)/2}
},  
\end{align}
and $R_{\rm dec}$ is the deceleration radius at which the swept-up CSM mass becomes comparable to the ejecta mass,
\begin{equation}
R_{\rm dec}\simeq
\left[
\frac{(3-w) M_{\rm ej}}{4\pi D}
\right]^{1 \over 3-w}. \label{eq:Rdec}
\end{equation}
Imposing continuity, the deceleration time $t_{\rm dec}$ is determined by the condition $R_{\rm dec} = R_{\rm s}(t_{\rm dec})$. 

Once $R_{\rm s}>R_{\rm dec}$, the system enters the blast-wave phase. In the adiabatic regime, the shock radius evolves as $R_{\rm s}\propto t^{\lambda}$ with $\lambda=2/(5-w)$, corresponding to the Sedov--Taylor solution \cite{Ostriker:1988,Suzuki:2020qui}. If radiative cooling is efficient, the evolution instead follows a shallower, momentum-conserving scaling with $\lambda=1/(4-w)$ \cite{Ostriker:1988,Moriya:2013hka}. As shown below, for the parameter space explored, the forward shock is generally expected to be radiative \cite{Moriya:2013hka}. We therefore adopt the radiative blast-wave evolution throughout this study. We have also verified that adopting the adiabatic evolution instead leads to a negligible impact on the resulting neutrino signal.

The shock velocity $v_{\rm s}$ is obtained by differentiating $R_{\rm s}$ with respect to $t$. Because $R_{\rm s}$ follows different temporal scalings below and above $R_{\rm dec}$, a direct differentiation of Eq.~\eqref{eq:Rs} would introduce an artificial discontinuity in $V_{\rm s}$ at $t=t_{\rm dec}$. To avoid this unphysical behavior, we adopt an empirical interpolation that smoothly connects the two asymptotic regimes,
\begin{equation}
V_{\rm s}(t)=\overline{V}_{\rm s}(t)
\left(1+\frac{\eta t}{t_{\rm dec}}\right)^{-\Delta},
\label{eq:vs_interp}
\end{equation}
where
\begin{align}
{\overline{V}}_{\rm s}(t) = \dfrac{n-3}{n-w}\left[
\dfrac{(3-w)(4-w)g_n}{(n-4)(n-3)D} \right]^{\frac{1}{n-w}} t^{\frac{w-3}{n-w}},
\end{align}
corresponds to the self-similar solution prior to deceleration. We choose $\Delta = {n-3 \over n-w}-\lambda$ and adopt $\eta \approx 4$, such that the late-time behavior at $t\gg t_{\rm dec}$ smoothly approaches, to good accuracy, the shock velocity obtained by differentiating $R_{\rm s}$ in Eq.~\eqref{eq:Rs}.

The shock transitions to a collisionless regime above the breakout radius $R_{\rm bo}$, where the optical depth satisfies
\begin{align}
\tau(R_{\rm bo}) \equiv \kappa_{\rm es} \int_{R_{\rm bo}}^{R_{\rm csm}} \rho_{\rm csm}(r) dr = {c\over V_{\rm s}(R_{\rm bo})},\label{eq:Rbo}
\end{align}
with $\kappa_{\rm es} \sim 0.34~{\rm cm^2~g^{-1}}$ being the electron-scattering opacity for a CSM of solar composition \cite{Pitik:2023vcg}, and $c$ the speed of light. If $\tau(R_\star)\le 1$ at the stellar surface, then the breakout radius coincides with the stellar radius, $R_{\rm bo}=R_\star$. Once the shock becomes collisionless, proton acceleration via the Fermi mechanism becomes efficient. The breakout radius therefore marks the onset of neutrino emission at time $t_{\rm bo}$, defined by $R_{\rm s}(t_{\rm bo})=R_{\rm bo}$, and the emission continues until the shock reaches the outer edge of the CSM at time $t_{\rm csm}$, where $R_{\rm s}(t_{\rm csm})=R_{\rm csm}$.  

The evolution of accelerated protons between $R_{\rm bo}$ and $R_{\rm csm}$ is described by \cite{Petropoulou:2017ymv,Pitik:2023vcg}
\begin{align}
{\partial N_p(\gamma_p, r) \over \partial r} & -{\partial \over \partial \gamma_p}\Big[{\gamma_p \over r} N_p(\gamma_p, r) \Big] \nonumber \\
& + {N_p(\gamma_p, r)\over V_{\rm s}(r)t_{pp}(r)} = Q_p(\gamma_p, r), \label{eq:Np_evol}
\end{align}
where $N_p(\gamma_p, r)$ denotes the total number of protons in the thin shocked shell with Lorentz factors between $\gamma_p$ and $\gamma_p+d\gamma_p$ at radius $r$. The second term accounts for adiabatic energy losses due to shell expansion, while the third term represents proton losses from inelastic $pp$ collision, characterized by a timescale $t_{pp}$. As indicated in the equation, we assume a constant $pp$ cross section, such that $t_{pp}$ is independent of $\gamma_p$. The source term $Q_p$ term denotes the injection rate of shock-accelerated protons as the forward shock sweeps up the CSM. The formal solution of Eq.~\eqref{eq:Np_evol} can be written in integral form:
\begin{align}
N_p(\gamma_p, r) = \int_{R_{\rm bo}}^{r} \frac{s}{r} e^{-\int_{s}^{r} [V_{\rm s}(u) t_{pp}(u)]^{-1} du} Q_p\left( \frac{s \gamma_p}{r}, s \right) ds.
\label{eq:Np_sol}
\end{align}

Assuming that a fraction $\epsilon_p$ of the dissipated kinetic energy is converted into the acceleration of protons, and adopting a power-law spectrum with index $p=2$, the injection rate $Q_p$ (in unit of ${\rm cm}^{-1}$) can be expressed as \cite{Pitik:2023vcg,Lu:2025jks}:
\begin{align}
Q_p(\gamma_p, r) = {9\pi \epsilon_p r^2 \rho_{\rm csm}(r) V^2_{\rm s}(r) \over 8 \ln(\gamma_{p, \rm max}/\gamma_{p, \rm min})m_pc^2} \gamma_p^{-2},
\end{align}
where $\gamma_{p, \rm min}$ and $\gamma_{p, \rm max}$ are the minimal and maximal Lorentz factors of the accelerated protons at each radius. We take $\gamma_{p, \rm min}=1$ and determine $\gamma_{p, \max}$ by comparing the proton acceleration and cooling timescales at each radius. The proton acceleration timescale is $t_{\rm acc} \sim 6\gamma_p m_p c^3/(eBV_{\rm s}^2)$, where $B=(9\pi\epsilon_BV_{\rm s}^2\rho_{\rm csm})^{1/2}$ and $\epsilon_B$ is the fraction of energy dissipated into magnetic field. The total cooling timescale can be written as $t_{\rm tot}^{-1} = t_{pp}^{-1} + {\rm max}[t_{\rm dyn}^{-1}, t_{\rm cool}^{-1}]$, where the cooling due to $pp$ collision can be approximated as $t_{pp} \simeq (4\kappa_{pp} \sigma_{pp}c\rho_{\rm csm}/m_p)^{-1}$ with $\kappa_{pp} \simeq 1$ and $\sigma_{pp}\simeq 3\times 10^{-26}~{\rm cm}^{-2}$, the dynamic timescale is $t_{\rm dyn} \sim r/V_{\rm s}$, and $t_{\rm cool}$ denotes the radiative cooling timescale [see Eq. (10) of \cite{Pitik:2023vcg}]. As the shock operates at large radii, synchrotron cooling can be safely neglected.

\begin{figure}[htbp]
\centering
\includegraphics[width=\columnwidth]{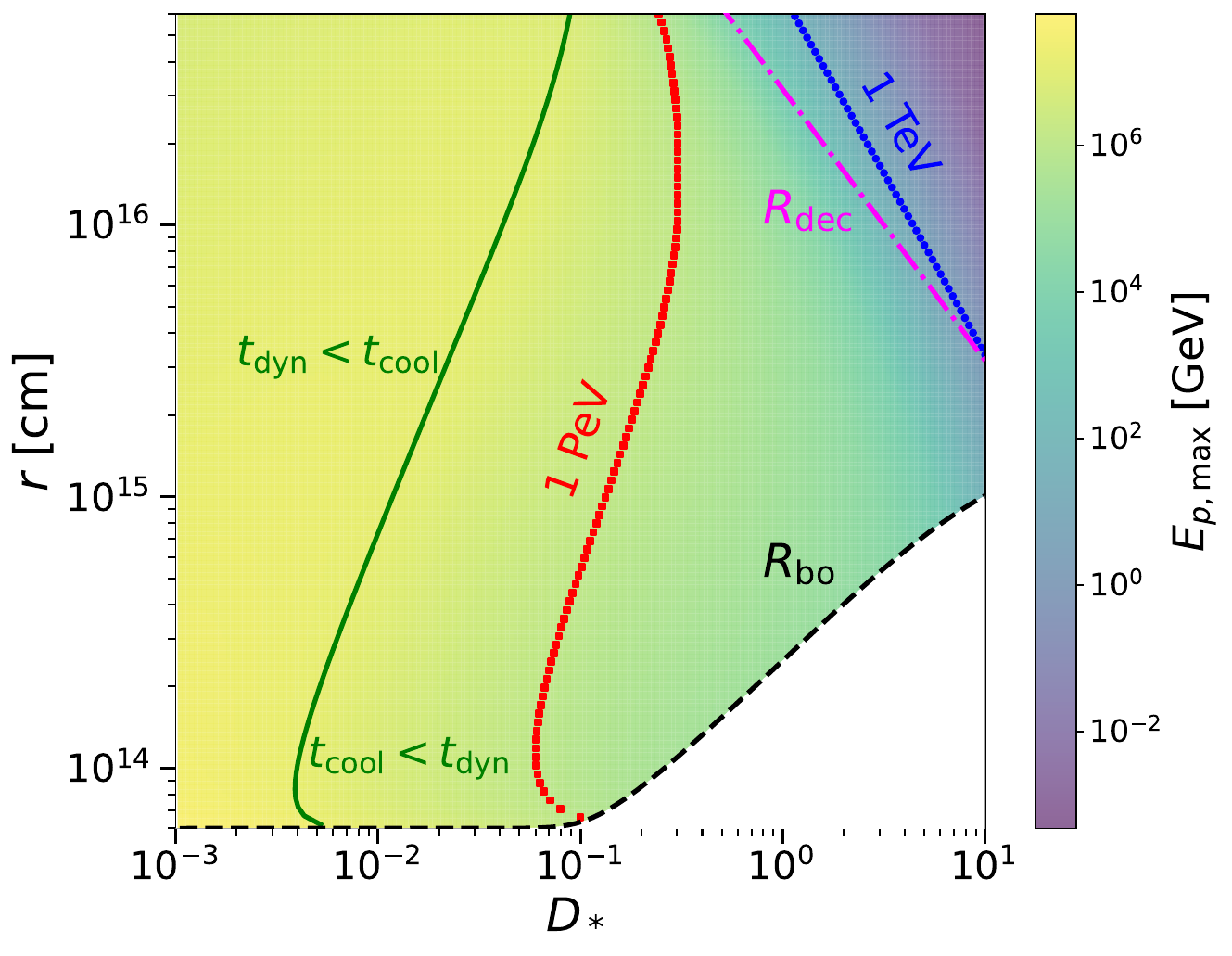}
\caption{Maximal attainable proton energy $E_{p, \max}$ as a function of $D_*$ and $r$. The contours corresponding to $E_{p, \max}=1\,\mathrm{TeV}$ and $E_{p, \max}=1\,\mathrm{PeV}$ are shown by the blue and red lines, respectively. The green line indicates the condition $t_{\rm dyn}=t_{\rm cool}$, with $t_{\rm dyn}<t_{\rm cool}$ ($t_{\rm cool}<t_{\rm dyn}$) to the left (right) of the line. Also shown are the breakout radius $R_{\rm bo}$ (black dashed) and the deceleration radius $R_{\rm dec}$ (magenta dash--dotted), both varying with $D_*$. Note that $R_{\rm csm}=10^{16}$~cm is adopted when computing $R_{\rm bo}$ using Eq.~\eqref{eq:Rbo}.}
\label{fig:DR-diagram}
\end{figure}

Figure~\ref{fig:DR-diagram} shows the maximal attainable proton energy $E_{p,\max}\equiv \gamma_{p,\max} m_p c^2$ as a function of $D_*$ and radius $r$, adopting $\epsilon_B=0.1$. In realistic situations, the radius should be bounded by $R_{\rm csm}$; here, for illustration, we assume that the CSM extends to all radii shown in the figure with a wind density profile. As $D_*$ increases, both $t_{\rm cool}$ and $t_{pp}$ decrease, while $t_{\rm dyn}$ increases slowly as $t_{\rm dyn}\propto D_*^{1/8}$ below $R_{\rm dec}$. Overall, proton cooling is dominated by dynamical expansion, radiative losses, and $pp$ interactions at small, intermediate, and large $D_*$, respectively. The contour corresponding to $t_{\rm dyn}=t_{\rm cool}$ is explicitly shown by the green line, to the left (right) of which $t_{\rm dyn}$ ($t_{\rm cool}$) dominates. At large $D_*$, the total cooling rate increases with $D_*$, thereby limiting the attainable value of $E_{p,\max}$. This is clearly illustrated by the contours of $E_{p,\max}=1\,\mathrm{TeV}$ (blue) and $E_{p,\max}=1\,\mathrm{PeV}$ (red). For $E_{p,\max}=1\,\mathrm{TeV}$ ($1\,\mathrm{PeV}$), the maximal energy is primarily determined by the competition between $t_{pp}$ ($t_{\rm cool}$) and the acceleration timescale $t_{\rm acc}$.

We also show the breakout radius $R_{\rm bo}$ (black dashed) and the deceleration radius $R_{\rm dec}$ (magenta dash--dotted) as functions of $D_*$, calculated using Eqs.~\eqref{eq:Rbo} and Eq.~\eqref{eq:Rdec}, respectively. In determining $R_{\rm bo}$, the upper bound in Eq.~\eqref{eq:Rbo} is fixed to $R_{\rm csm}=10^{16}\,\mathrm{cm}$. As seen, $t_{\rm cool}<t_{\rm dyn}$ at $R\gtrsim R_{\rm dec}$, indicating that the shocked shell is likely radiative during the blast-wave phase. Note that the blast-wave phase occupies only a limited region of the $D_*$--$r$ parameter space and therefore plays a minor role in this study.

Once $N_p(\gamma_p, r)$ is obtained, the all-flavor, energy-differential neutrino emissivity (in units of $\rm GeV^{-1}~s^{-1}$) from $pp$ collisions can be expressed as
\begin{align}
I_\nu(E_\nu, t) \equiv {dN_\nu \over dE_\nu dt} & = \int_{\gamma_{p, \rm min}}^{\gamma_{p, \rm max}} N_p(\gamma_p, R_{\rm s}(t)) 
\frac{4\rho_{\rm csm}(r)\sigma_{pp}c}{m_p}
\nonumber \\
& \times Y_\nu(E_p=\gamma_p m_pc^2, E_\nu)   
d\gamma_p,\label{eq:Inu}
\end{align}
where $Y_\nu(E_p, E_\nu)$ is the averaged neutrino spectrum produced per $pp$ collision. To obtain $Y_\nu$, we employ PYTHIA 8.3 \cite{Bierlich:2022pfr} to simulate the production of $\pi^\pm$ and $K^\pm$, and to compute the resulting HE neutrino spectra by accounting for the decays of $\pi^\pm$ and $K^\pm$, and the subsequent decay of $\mu^\pm$, following the procedures described in Refs. \cite{Guo:2022zyl,Guo:2025wkt,Xiao:2025yxc}.The cooling of mesons can be safely neglected under the conditions considered here. The total neutrino spectrum emitted over a given time interval $[t_1, t_2]$ is then obtained by integrating the time-dependent emissivity over that period.  

As commonly adopted in the literature \cite{Murase:2017pfe}, the neutrino flux can be estimated without explicitly solving the proton transport equation. In this simplified approach, a fraction $f_{pp}$ of the injected protons is assumed to interact inelastically with the ambient medium to produce HE neutrinos through $pp$ collisions. The resulting neutrino emissivity can then be expressed as
\begin{align}
I_\nu^0(E_\nu, t) \simeq & \int d\gamma_p Q_p(\gamma_p, R_{\rm s}(t)) f_{pp}(t) Y_\nu(E_p, E_\nu),\label{eq:Inu0}
\end{align}
where $E_p=\gamma_p m_p c^2$ and the effective interaction probability is approximated as $f_{pp}(t) \sim \min[t_{\rm dyn}/t_{pp}, 1] \sim \min[\sigma_{pp}(4\rho_{\rm csm}/m_p)R_{\rm s}(t)c/V_{\rm s}(t), 1]$.

\begin{figure*}[htbp]
\centering
\includegraphics[width=\columnwidth]{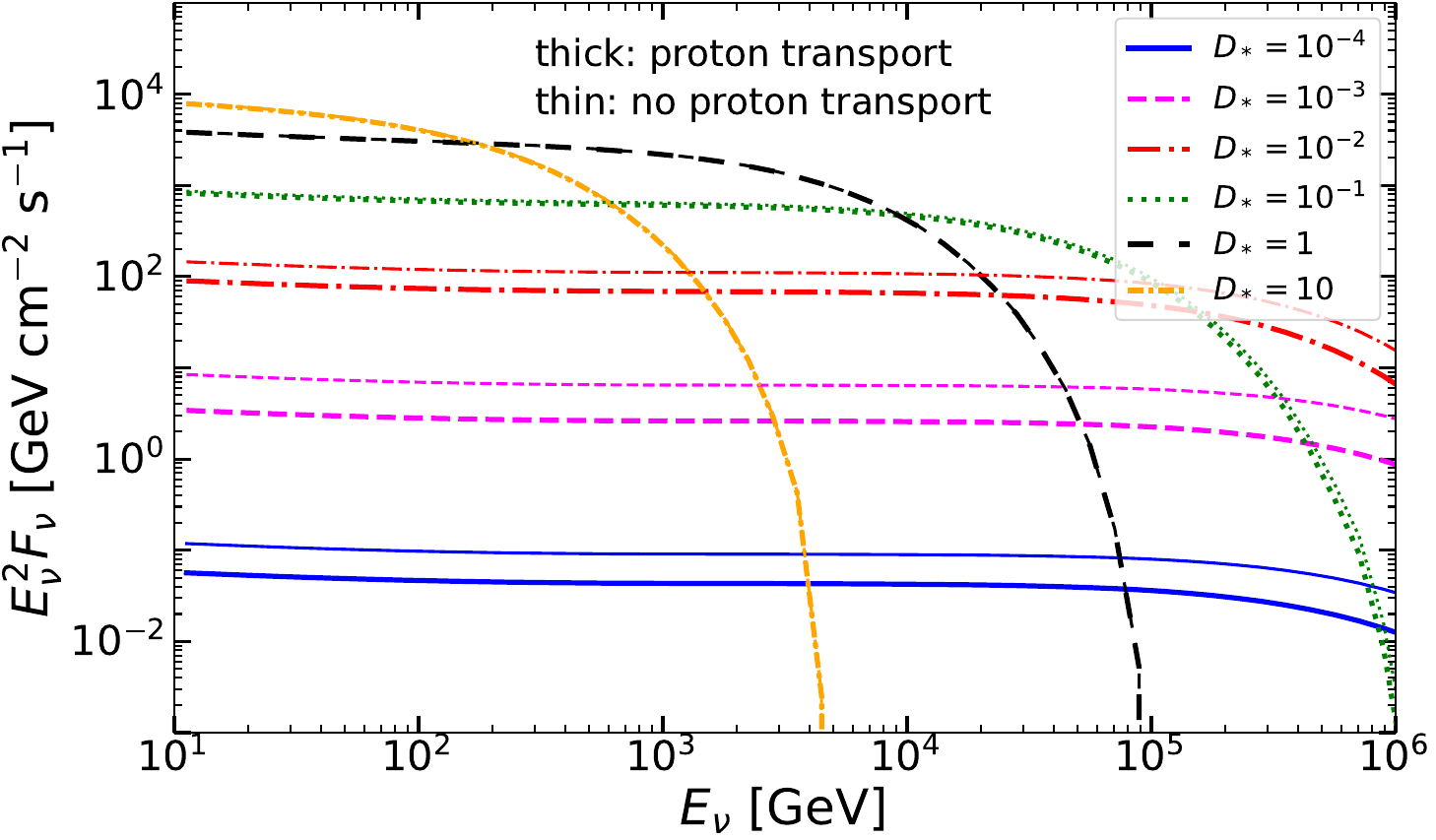}
\includegraphics[width=\columnwidth]{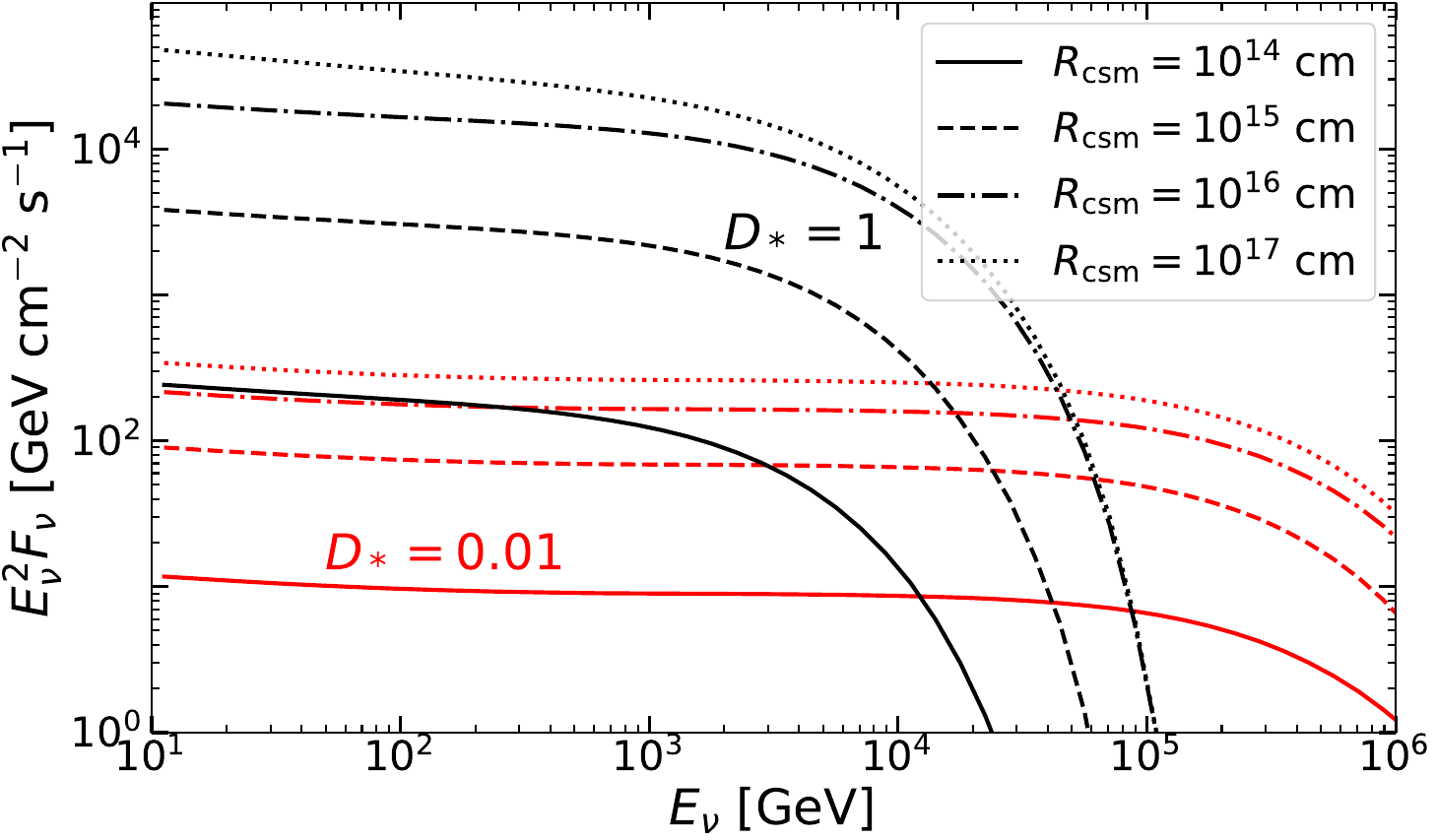}
\\
\includegraphics[width=\columnwidth]{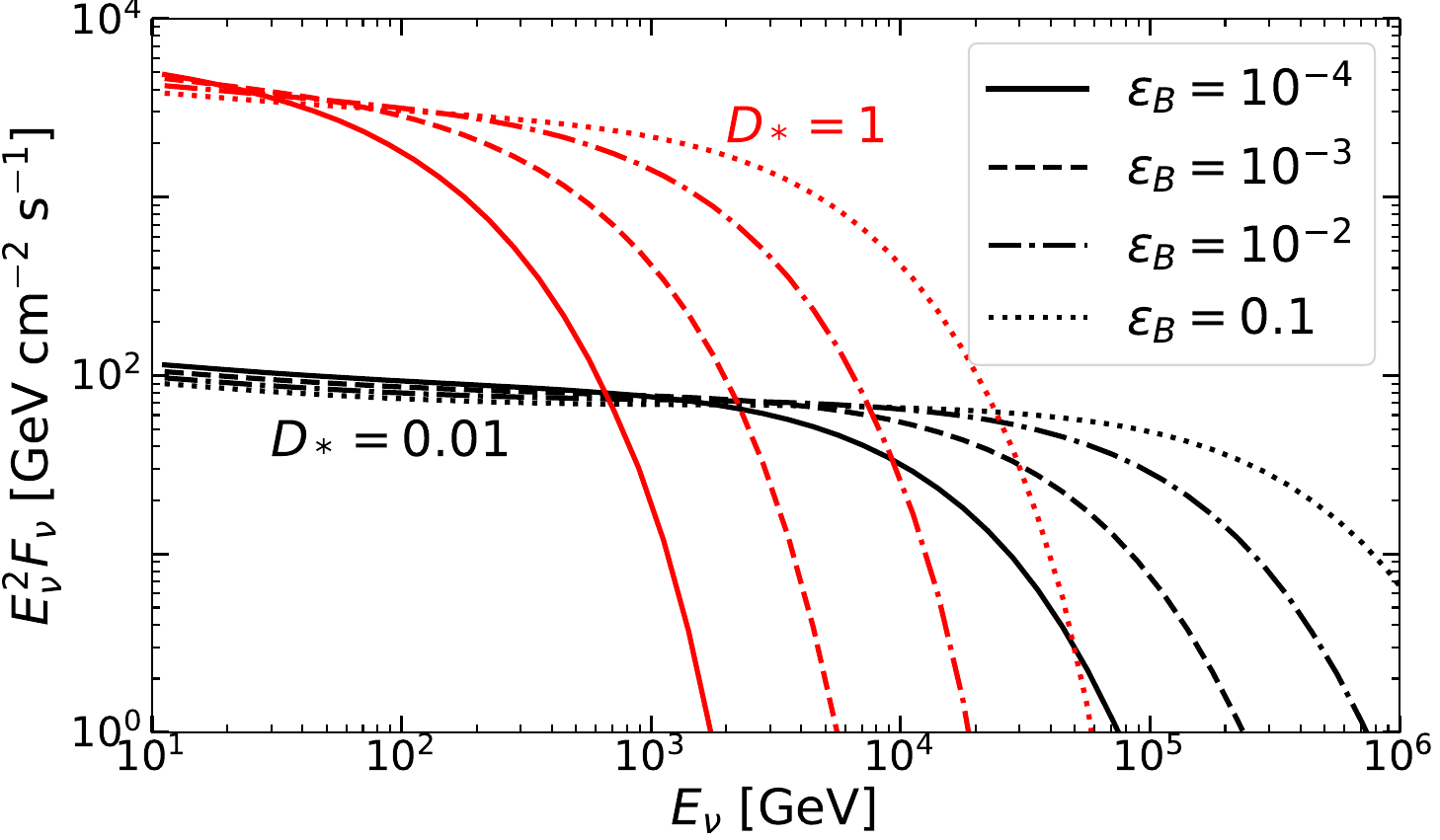}
\includegraphics[width=\columnwidth]{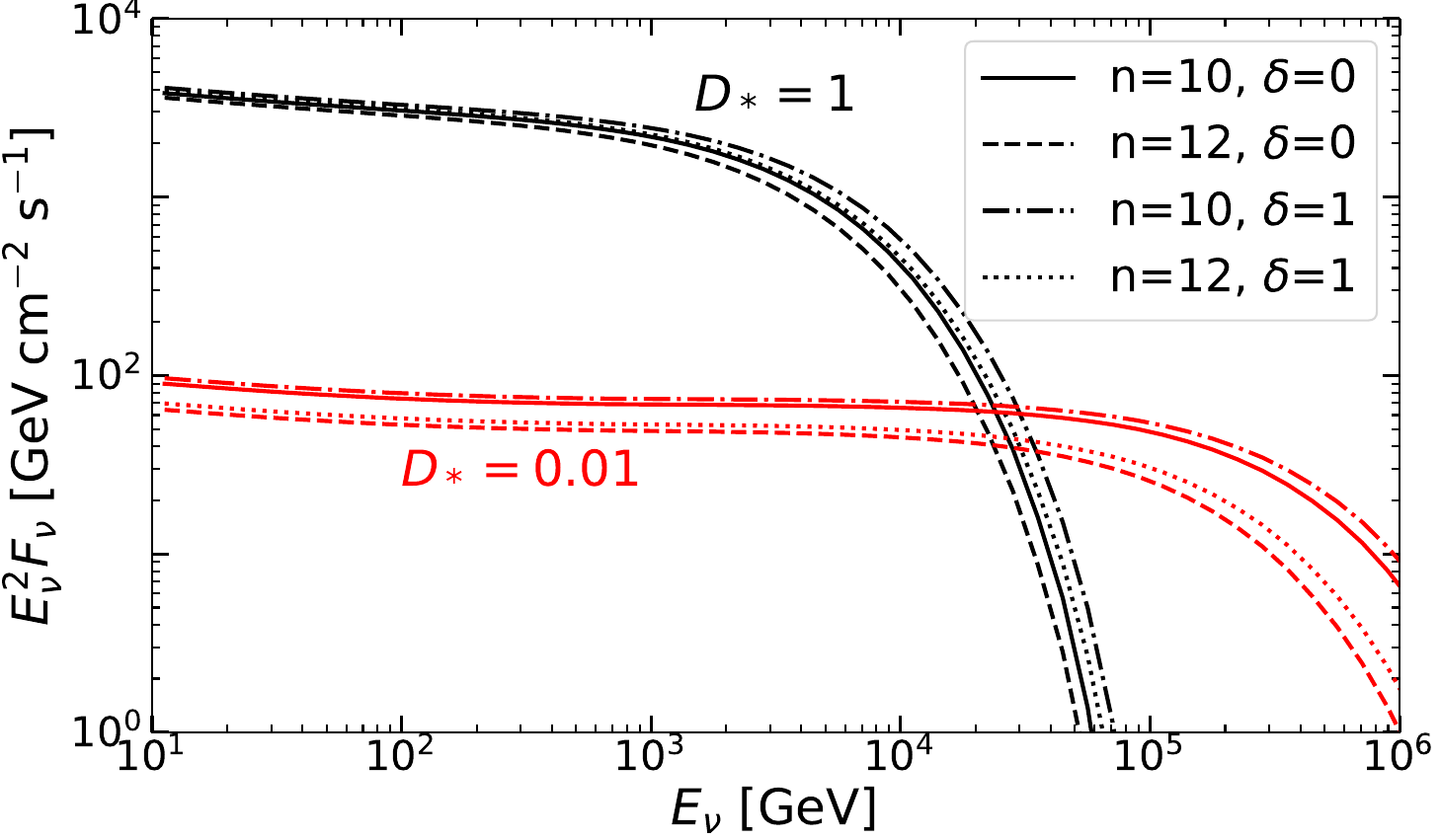}
\caption{Expected all-flavor neutrino fluxes from a nearby CCSN at $L=10$~kpc using different values of $D_*$ (upper left), $R_{\rm csm}$ (upper right), $\epsilon_B$ (lower left), and $n$ and $\delta$ (lower right). In the upper panel, neutrino fluxes calculated with (thick) or without (thin) solving the proton transport equation are also compared. We adopt $D_* = 0.01$ or 1, $R_{\rm csm}=10^{15}$~cm, $\epsilon_B=0.1$, and $\epsilon_p=0.1$, unless otherwise specified.}
\label{fig:fluxes}
\end{figure*}

\begin{figure}[htbp]
\centering
\includegraphics[width=\columnwidth]{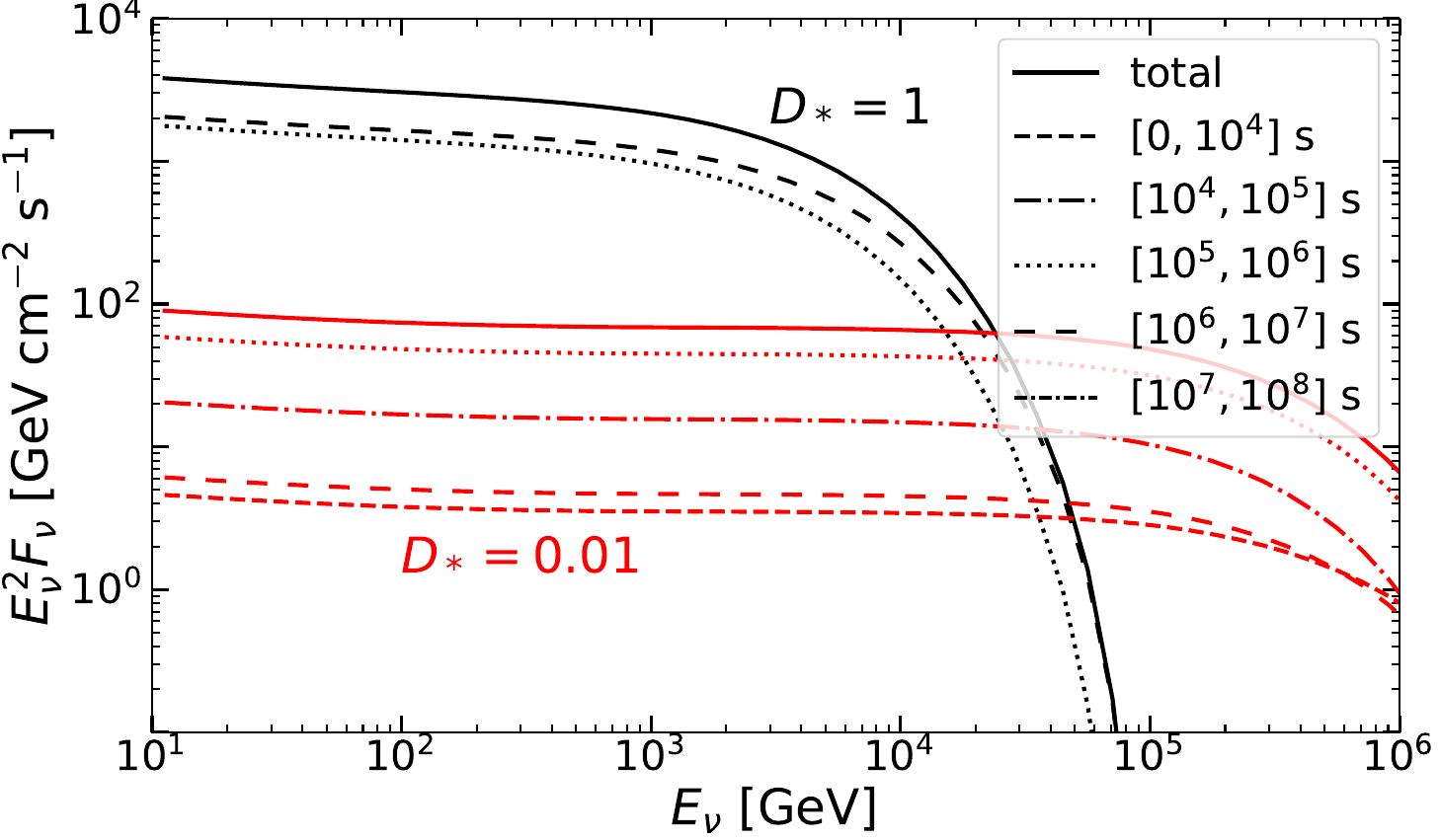}
\caption{Expected all-flavor neutrino fluxes from a nearby SN at $L=10$~kpc with $D_*=0.01$ and 1. Five different time windows after the explosion have been considered. The same parameter values as in Fig.~\ref{fig:fluxes} are taken.}
\label{fig:flux_timewindow}
\end{figure}

\subsection{the resulting neutrino fluxes}

As seen, the neutrino fluence from the interaction of the SN ejecta with the CSM is governed by a set of physical parameters. To illustrate the results in this section, we adopt the following fiducial values unless otherwise specified: $D_* = 0.01$ or 1, $R_{\rm csm}=10^{15}$~cm, $\epsilon_B=0.1$, and $\epsilon_p=0.1$.

The resulting neutrino flux from a nearby SN located at 10 kpc and its dependence on a few relevant parameters, including $D_*$, $R_{\rm csm}$, $\epsilon_B$, $n$ and $\delta$, are presented in Fig.~\ref{fig:fluxes}. The upper left panel shows the significant influence of $D_*$ on the neutrino flux. A wide range of $D_*$ values, $D_*=10^{-4}$, $10^{-3}$, $10^{-2}$, 0.1, 1, 10, is considered. As expected, lower $D_*$ values result in a reduced overall flux normalization but a higher cutoff energy in the neutrino spectrum. This can be understood as follows: a lower CSM density reduces both the amount of swept-up CSM and the rate of $pp$ collisions, thereby decreasing the flux, but it also increases the $pp$ cooling timescale. Consequently, protons can be accelerated to higher energies before interacting, raising the maximum proton energy and shifting the neutrino spectrum cutoff to higher energies.

This panel also compares the neutrino fluxes obtained by solving the proton transport equation [thick lines, Eqs.~\eqref{eq:Np_evol} and \eqref{eq:Inu}] and those calculated without solving it [thin lines, Eq.~\eqref{eq:Inu0}]. As shown in the figure, explicitly including proton transport accounts for the cooling of HE protons toward lower energies, which suppresses their contribution to the HE neutrino flux. In contrast, the simplified treatment without proton transport assumes that the accelerated protons effectively interact with the ambient medium over a dynamical timescale $t_{\rm dyn}$, without properly taking into account the cooling of protons during their evolution. We find that including proton transport can reduce the neutrino flux by a factor of $\sim 2$. This effect becomes less pronounced at higher CSM densities (e.g., $D_*\gtrsim 0.1$), where the increased target density results in a shorter $pp$ collision timescale and adiabatic cooling becomes negligible.

The upper right panel compares the neutrino flux for different values of $R_{\rm csm}$ ($10^{14}$~cm, $10^{15}$~cm, $10^{16}$~cm, and $10^{17}$~cm), assuming $D_*=0.01$ and 1. A more extended CSM generally leads to a higher total neutrino flux compared to a compact configuration, owing to the larger total CSM mass swept up for $pp$ interactions, which sustains neutrino emission over a longer duration. Despite the fact that the $D_*$ values for both subclass can be comparable, the more compact CSM leads to lower neutrino fluxes from typical Type II SNe compared to Type IIn SNe, which possess more extended CSM. Note that the dependence on $R_{\rm csm}$ is not linear. As the shock propagates to larger radii, the decreasing CSM density reduces the $pp$ interaction optical depth, leading to a saturation effect that further increases in $R_{\rm csm}$ result in diminishing gains in the total integrated neutrino flux.

The dominant effect of $\epsilon_B$ is on the proton acceleration timescale, $t_{\rm acc}\propto E_p/B \propto E_p\epsilon_B^{-1/2}$. As all the relevant cooling time scales, $t_{\rm dyn}$, $t_{\rm cool}$, and $t_{pp}$, are independent of $E_p$, the balance between the cooling and acceleration timescales implies $E_{p, \max} \propto \epsilon_B^{1/2}$. As a result, the neutrino spectrum extends to higher energies with increasing $\epsilon_B$
(see the lower left panel of Fig.~\ref{fig:fluxes} for $\epsilon_B=10^{-4}$, $10^{-3}$, $10^{-2}$, and 0.1).

The parameters $n$ and $\delta$, which characterize the SN ejecta structure, have only a relatively minor impact on the neutrino flux, primarily affecting its overall normalization. As shown in the lower right panel of Fig.~\ref{fig:fluxes}, variations within typical ranges ($n = 10$--12 and $\delta = 0$--1) lead to changes of up to $\pm 20$--30\%. In the following studies, we simply fix $n=10$ and $\delta=0$. Although not shown, the neutrino flux scales linearly with $\epsilon_p$. Since the effects of $n$ and $\delta$ effectively act as a normalization uncertainty, they can be absorbed into $\epsilon_p$, whose uncertainty will be taken into account.

It is worthy noting that different combinations of parameters can produce similar neutrino fluxes. For example, both $D_*$ and $\epsilon_B$ influence the neutrino spectrum: $\epsilon_B$ primarily affects the maximum proton energy (and thus the spectral cutoff), while $D_*$ impacts both the cutoff and the overall flux normalization. Different combinations of $D_*$ and $\epsilon_B$, with appropriate tuning of $\epsilon_p$, can therefore produce similar neutrino signals. The temporal evolution of the neutrino signal can help break these parameter degeneracies, since $D_*$ effectively controls the emission duration. Fig.~\ref{fig:flux_timewindow} shows the expected neutrino flux within successive time windows: $[0, 10^4]$~s, $[10^4, 10^5]$~s, $[10^5, 10^6]$~s, $[10^6, 10^7]$~s, and $[10^7, 10^8]$~s. As more CSM is swept up at later times, these windows generally contribute more strongly to the total signal, particularly at lower energies. However, for low $D_*$ values (e.g., $D_* = 0.01$), the CSM density at large radii becomes sufficiently low that the $pp$ interaction probability is significantly suppressed, causing the windows of $[10^{6}, 10^{7}]$ s and $[10^{7}, 10^{8}]$ s to contribute less than the earlier interval. For large $D_*$ (e.g., $D_* = 1$), the shock breakouts at large $R_{\rm bo}$ (see Fig.~\ref{fig:DR-diagram}), and as a result, neutrino emission begins at later times.

\section{Detection Prospects for a Nearby SN and the Diffuse Flux}
\label{sec:detection}

The mass-loss rate distributions and neutrino fluxes derived in Sec.~\ref{sec:basic} serve as the astrophysical inputs for evaluating the detectability of these neutrino signals. Based on these results, we first examine the detection prospects of HE neutrinos from a nearby SN, adopting representative values of $D_*$ and $R_{\rm csm}$ for Type~II and Type~IIn SNe. We then compute the resulting diffuse neutrino flux from these two subclasses, accounting for the distribution of $D_*$.

Unless otherwise specified, we adopt $D_* = 0.01$ and $R_{\rm csm} = 4 \times 10^{14}\,\mathrm{cm}$ for Type~II SNe \cite{Yaron:2017,Morozova:2017,Hinds:2025qia}, and $D_* = 1$ and $R_{\rm csm} = 10^{16}\,\mathrm{cm}$ for Type~IIn SNe \cite{Salmaso:2024jry}, both located at a distances of $10\,\mathrm{kpc}$. We further set $\epsilon_p = 0.1$ and consider different values of $\epsilon_B$ raging from $10^{-4}$--0.1. These benchmark parameter values are summarized in Table~\ref{tab:para}. When estimating the cumulative diffuse flux, we employ the observationally inferred $D_*$ distributions, while adopting the benchmark values of $R_{\rm csm}$ for each subclass and $\epsilon_p$.

\begin{table}[ht]
\centering
\renewcommand{\arraystretch}{1.5}
\caption{Benchmark parameter values adopted for Type~II and Type~IIn SNe in our study of the detection prospects for neutrino signals from a nearby SN. We fix the remaining parameters to $R_\ast = 6 \times 10^{13}\,\mathrm{cm}$, $M_{\rm ej} = 10\,M_\odot$, $\mathcal{E}_{\rm ej} = 10^{51}\,\mathrm{erg}$, $n = 10$, and $\delta = 0$, as mentioned in Sec.~\ref{sec:basic}.
}
\begin{tabular}{|w{c}{1.5cm}|w{c}{1.cm}|w{c}{2.0cm}|w{c}{2.cm}|w{c}{1.0cm}|}
\hline 
 & $D_*$ & $R_{\rm csm}$ & $\epsilon_B$ & $\epsilon_p$ \\
\hline
Type II & 0.01 & $4\times 10^{14}$~cm & $10^{-4}$, 0.1 & 0.1 \\
\hline
Type IIn & 1 & $10^{16}$~cm & $10^{-4}$, 0.1 & 0.1 \\
\hline
\end{tabular}
\label{tab:para}
\end{table}

\subsection{Detecting HE neutrinos from a nearby SN} \label{sec:detection_single}
 
We focus on IceCube \cite{IceCube:2016zyt} to study the neutrino signal, although the approach can be readily extended to other detectors such as ANTARES \cite{ANTARES:2011}, KM3NeT \cite{KM3Net:2016zxf}, and planned next-generation telescopes \cite{IceCube-Gen2:2020qha,P-ONE:2020ljt,TRIDENT:2022hql}. Two primary event topologies are commonly used in searches for astrophysical neutrinos: track events, which are mainly produced by charged-current (CC) interactions of $\nu_\mu$ and $\bar\nu_\mu$ (with a subdominant contribution from CC interactions of $\nu_\tau$ and $\bar\nu_\tau$ through tau decays), and cascade events, arising from CC interactions of electron and tau neutrinos as well as neutral-current interactions of all flavors. Owing to their substantially larger effective area and superior angular resolution, track events play a key role in searches for neutrinos from astrophysical sources (see, e.g., \cite{IceCube:2015qii,IceCube:170922A,TXS:2018,NGC:2022,KM3NeT:2025npi,Abbasi:2025tas}). Note that atmospheric muons constitute a major background for downgoing track events, while for upgoing tracks, this contamination is strongly suppressed by the shielding effect of the Earth. To detect HE neutrinos from the Southern Hemisphere with IceCube, cascade events--especially starting cascades--are better suited, as they efficiently reject the atmospheric muon background \cite{IceCube:2019lzm,IceCube:2023ame,IceCube:2024fxo}. For illustration, we restrict our analysis to upgoing track events, corresponding to sources located in the Northern Hemisphere. We note, however, that the inclusion of starting cascade events could yield comparable detection prospects for Northern Hemisphere sources, as demonstrated in Ref.~\cite{Xiao:2025yxc}.    

\begin{figure*}[htbp]
\centering
\includegraphics[width=\columnwidth]{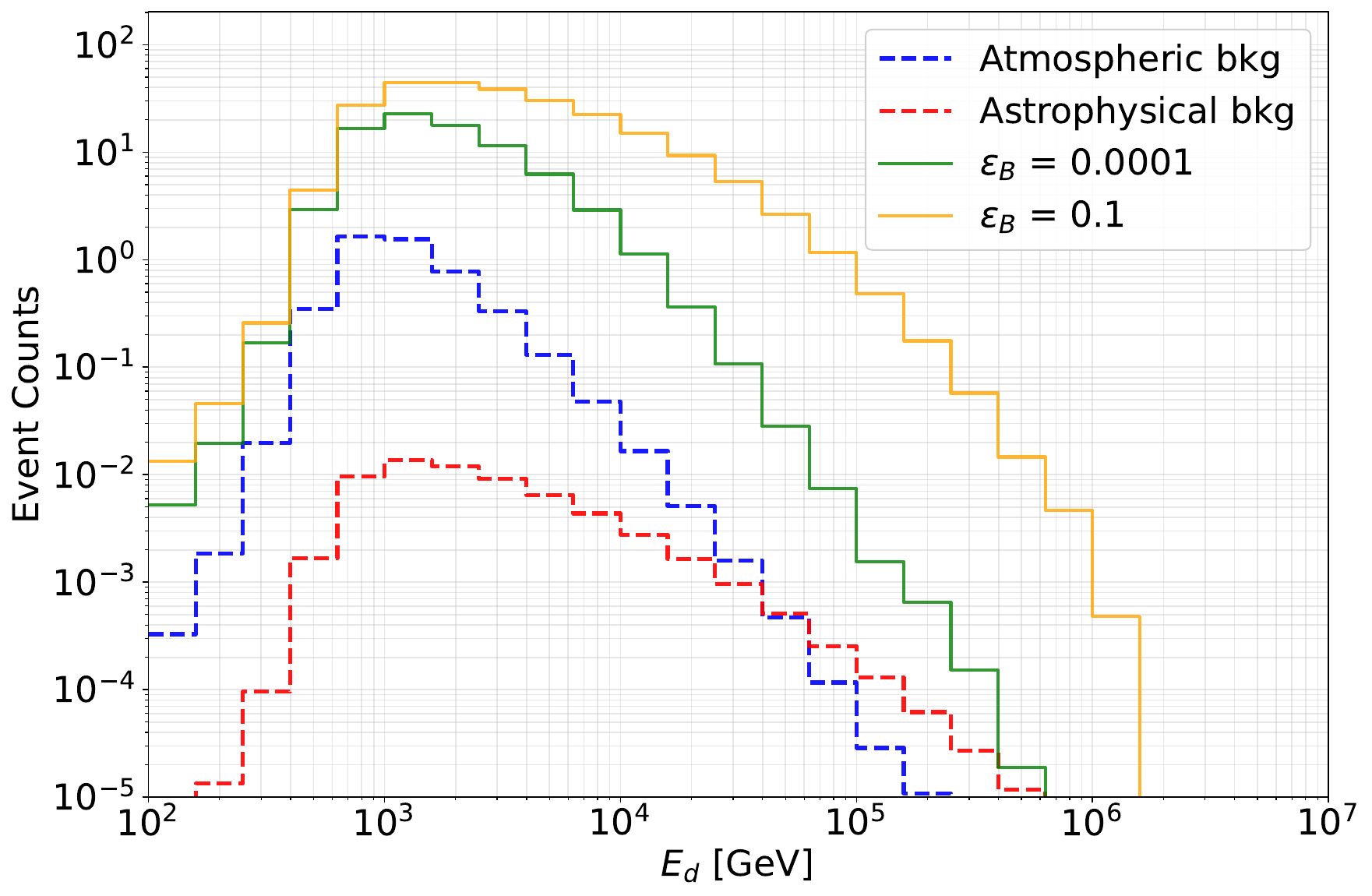}
\includegraphics[width=\columnwidth]{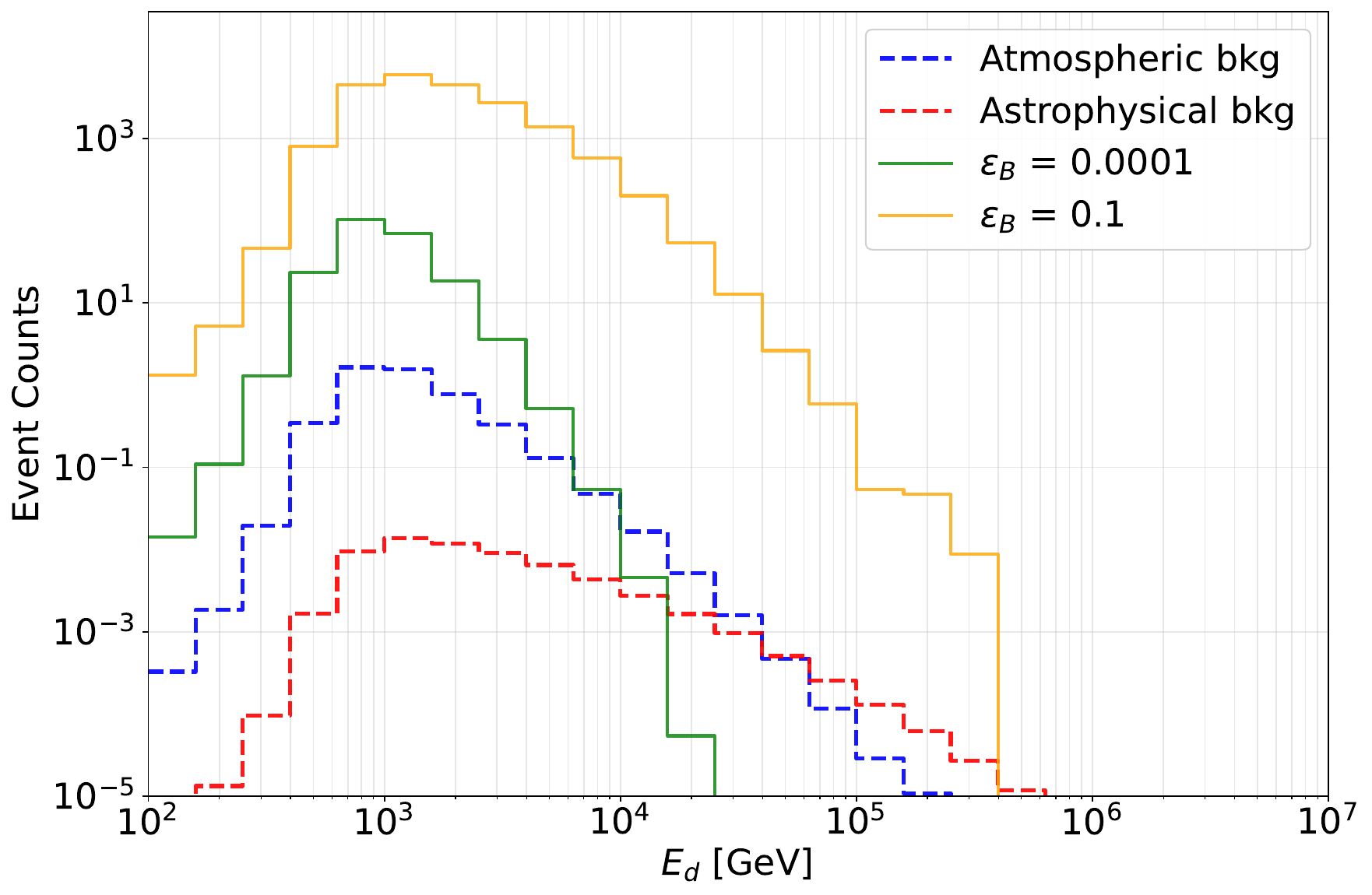}
\caption{Expected track spectra at IceCube from a Type~II SN (left panel) with $D_* = 0.01$ and $R_{\rm csm} = 4 \times 10^{14}\,\mathrm{cm}$, and from a Type~IIn SN (right panel) with $D_* = 1$ and $R_{\rm csm} = 10^{16}\,\mathrm{cm}$, both located at a distance of $L = 10\,\mathrm{kpc}$. Two values of $\epsilon_B = 10^{-4}$ and 0.1 are adopted. For parameters not explicitly specified, the benchmark values listed in Table~\ref{tab:para} are adopted.
}
\label{fig:observed}
\end{figure*}

To quantitatively assess the discovery potential of HE neutrinos from nearby SNe, we perform a binned likelihood analysis similar to that adopted in Ref.~\cite{Xiao:2025yxc}. We always assume that the source distance $L$ is well known\footnote{The uncertainty in $L$ can be absorbed into $\epsilon_p$.}, and that a given SN ejecta-CSM interaction model can be characterized by the parameter set 
\begin{align}
\boldsymbol{\lambda} \equiv \{ D_*,\; R_{\rm csm},\; \epsilon_B,\; \epsilon_p \}.
\label{eq:lambda}
\end{align}
The expected number of track events induced by neutrinos in each time and energy bin is then given by
\begin{align}
s_{ij}(\boldsymbol{\lambda}) = \sum_{f=\nu_\mu,\bar\nu_\mu,\nu_\tau,\bar\nu_\tau} \int_{t_i^{\rm low}}^{t_i^{\rm up}} dt \int dE_\nu F_f(E_\nu, t) A_{f,j}(E_\nu,\theta), \label{eq:s_ij}
\end{align}
where $t_i^{\rm low, up}$ are the lower and upper bounds of the $i$-th time bin, $F_f(E_\nu, t)$ (in unit of $\rm GeV^{-1}~cm^{-2}~s^{-1}$) is the time-dependent neutrino flux at Earth for flavor $f$, and $A_{f,j}(E_\nu, \theta)$ is the IceCube effective area for a neutrino of flavor $f$ to produce a track event with deposited energy $E_{\rm d}$ in the $j$-th energy bin, which depends on neutrino energy $E_\nu$ and zenith angle $\theta$. For simplicity, we assume that the neutrino flux is equally distributed between neutrinos and antineutrinos, and among all flavors as a result of neutrino oscillations. For upgoing tracks, we adopt the effective areas from~\cite{IceCube:2015qii,track-effetive-area}. We take five post-explosion time bins as in Fig.~\ref{fig:flux_timewindow}: $[0, 10^4]$~s, $[10^4, 10^5]$~s, $[10^5, 10^6]$~s, $[10^6, 10^7]$~s, and $[10^7, 10^8]$~s, and use the same energy bins for $E_{\rm d}$ as provided in the effective-area tables. 

Figure~\ref{fig:observed} shows the expected track spectra at IceCube produced by HE neutrinos from a Type II (left panel) and a Type IIn (right panel) SN located at 10 kpc, adopting the benchmark parameter values listed in Table~\ref{tab:para}. To illustrate the impact of the magnetic-field energy fraction, we consider $\epsilon_B = 10^{-4}$ and 0.1. In all cases, the deposited-energy spectra peak at $E_{\rm d} \sim 1$ TeV, while the suppression at higher energies is primarily caused by the absorption of HE neutrinos in the Earth. The higher event rates predicted for Type IIn SNe or for larger $\epsilon_B$ reflect the dependence of the neutrino flux on $D_*$, $R_{\rm csm}$, $\epsilon_B$, as discussed in Sec.~\ref{sec:basic}.

For comparison, we also show the expected background contributions from atmospheric neutrinos and diffuse astrophysical neutrinos, which constitute the dominant background for upgoing track events. The atmospheric component is computed using MCEq \cite{MCEq}, while the diffuse astrophysical neutrino flux is modeled by a single power-law spectrum based on the IceCube 9.5-year track data \cite{Abbasi:2021qfz}. The background rate is evaluated within a $1^\circ$ angular search window around the SN direction, corresponding to a solid angle $\Delta \Omega \simeq 0.001$. With this selection, the total background amounts to $\sim$ 3 events above 0.1 TeV, concentrated mainly at TeV energies. For a typical Galactic Type II SN, the predicted signal significantly exceeds the background over the relevant energy range, suggesting that backgrounds play only a minor role.

The likelihood function incorporating contributions from both signal and background is defined as
\begin{align}
\mathcal{L}(\{s_{ij}(\boldsymbol{\lambda})\}) &= \max_{\{f_b\}} \Big\{\prod_{ij} P(N_{ij}, s_{ij}(\boldsymbol{\lambda}) + (1+f_b) b_{ij}) \nonumber \\
&\times \exp\Big(-\frac{f_b^2}{2\sigma_b^2}\Big) \Big\},
\label{eq:likeli}
\end{align}
where $P(N, \mu)$ denotes the Poisson probability of observing $N$ events given an expectation value $\mu$, $N_{ij}$ is the total number of detected track events in the $i$-th time and $j$-th energy bin, and $s_{ij}$ is the predicted signal number depending on the SN interaction model [see Eq.~\eqref{eq:s_ij}],
and $b_{ij}$ is the expected background number. The parameter $f_b$ is introduced as a nuisance parameter to account for the overall normalization uncertainty of the background, with an associated variance $\sigma_b$. As indicated, the likelihood function in Eq.~\eqref{eq:likeli} is maximized with respect to $f_b$. We set $\sigma_b=0.2$, and note that the exact value does not affect the final results.

The discovery potential is quantified using the test statistic (TS), defined as the logarithmic likelihood ratio between the signal-plus-background and background-only hypotheses. Throughout this work, we compute the median discovery significance using the Asimov dataset \cite{Cowan:2010js}. For given SN model characterized by $\boldsymbol{\lambda}$, the observed counts $N_{ij}$ in Eq.~\eqref{eq:likeli} are set equal to the expected values under the signal-plus-background hypothesis: $N_{ij} = s_{ij}(\boldsymbol{\lambda}) + b_{ij}$. Using the definition of ${\cal L}$ in Eq.~\eqref{eq:likeli}, the TS is then given by
\begin{equation}
\mathrm{TS}(\boldsymbol{\lambda}) = 2 \big[ \ln \mathcal{L}(\{s_{ij}(\boldsymbol{\lambda})\}) - \ln \mathcal{L}(\{s_{ij}=0\}) \big]. \label{eq:TS}
\end{equation}

\begin{figure}[htbp]
\centering
\includegraphics[width=\columnwidth]{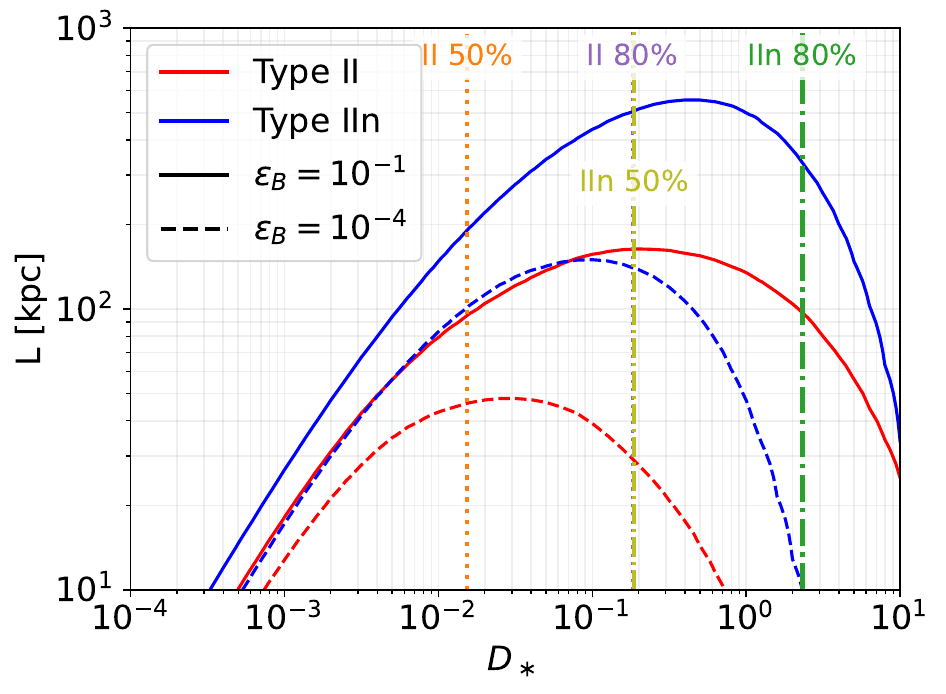}
\caption{Detection horizons for HE neutrinos from Type II (red) and Type IIn (blue) SNe as a function of $D_*$. We adopt a $3\sigma$ detection criterion, corresponding to ${\rm TS}=9$ [see Eq.~\eqref{eq:TS}]. Results are shown for $R_{\rm csm}=4 \times 10^{14}$~cm (Type II) and $R_{\rm csm}=10^{16}$~cm (Type IIn), assuming $\epsilon_B=10^{-4}$ (dashed) and 0.1 (solid). Vertical lines indicate the cumulative probabilities (50\% and 80\%) for Type~II and Type IIn SNe to have $D_*$ below a given value, based on the normalized mass-loss rate distributions shown in Fig.~\ref{fig:distribution}.}
\label{fig:horizon}
\end{figure}

Figure~\ref{fig:horizon} shows the resulting detection horizons at the $3\sigma$ confidence level (CL; ${\rm TS}=9$) as a function of $D_*$, assuming $R_{\rm csm}=4\times 10^{14}$~cm and $10^{16}$~cm, and $\epsilon_B=10^{-4}$ and 0.1. For regular Type II SNe, the detection horizon increases with $D_*$ for $D_* \lesssim 0.03$ ($\epsilon_B=10^{-4}$) or $D_* \lesssim 0.3$ ($\epsilon_B=0.1$), and then decreases at larger $D_*$, reflecting the dependence of neutrino flux on $D_*$ (see the upper left panel of Fig.~\ref{fig:fluxes}). In the optimal $D_*$ regime, the horizon extends to $\sim 0.05$--0.2~Mpc, depending on $\epsilon_B$. With a more extended CSM, the neutrino flux from Type IIn SNe exceeds that of normal Type II SNe for a given $D_*$. As a result, the detection horizon for Type IIn can reach $\sim 0.15$ Mpc for $\epsilon_B=10^{-4}$ and $\sim 0.6$ Mpc for $\epsilon_B=0.1$. The vertical lines indicate the cumulative probability contours (50\% and 80\%) that a Type~II (red) or Type IIn (blue) SN has $D_*$ below a given value, inferred from the observed mass-loss rate distributions. A substantial fraction of Type~II and Type IIn SNe lie within their respective optimal $D_*$ ranges, implying that a significant subset of nearby events occurs under favorable CSM conditions for HE neutrino detection. For typical Galactic Type II SNe ($L=10$ kpc) with $D_*$ in the range $10^{-3}$--1, HE neutrino detection at $\gtrsim 3\sigma$ CL can be achieved.

These results are broadly consistent with those of Ref.~\cite{Kheirandish:2022eox}, which similarly identified regular Type~II SNe located within $\sim 1$ Mpc with moderately dense winds as promising sources of HE neutrinos detectable by IceCube. Note that $\epsilon_p=0.1$ is adopted in Fig.~\ref{fig:DR-diagram}. As easily understood, the detection horizon scales as $\epsilon_p^{1/2}$ for fixed $D_*$ and $\epsilon_B$.

\subsection{Detection prospect for the diffuse neutrino flux}

The diffuse HE neutrino flux from Type II and Type IIn SNe is obtained by integrating the neutrino fluence from individual events over redshift,
\begin{align}
\label{eq:diff_flux}
\Phi_{\rm diff}(E_\nu)=\frac{c}{4\pi}\int_0^\infty & R_{\rm cc}(z) \Big\langle {dN_\nu \over dE_\nu}\!\left(E_\nu(1+z)\right)\Big\rangle \nonumber \\
& \times 
(1+z)\left|\frac{dt}{dz} \right| dz.
\end{align}
Here, $R_{\rm cc}(z)$ is the total CCSN rate~\cite{Ma:2025gbo},
and $dt/dz$ denotes the cosmic time-redshift relation, evaluated following Ref.~\cite{Xiao:2025yxc}. $\langle dN_\nu(E_\nu)/dE_\nu\rangle$ is the all-flavor time-integrated neutrino spectrum per CCSN (in unit of ${\rm GeV}^{-1}$), averaged over the distribution of $D_*$:
\begin{align}
\Big\langle & {dN_\nu(E_\nu)\over dE_\nu}\Big\rangle = \int \Big[\eta_{\rm II} f_{\rm II}(\log_{10}D_*) {dN^{(\rm II)}_\nu(E_\nu)\over dE_\nu} \nonumber \\ 
& + \eta_{\rm IIn} f_{\rm IIn}(\log_{10}D_*) {dN^{(\rm IIn)}_\nu(E_\nu)\over dE_\nu}\Big] d\log_{10}D_*,
\label{eq:fluence_avg}
\end{align}
with $dN^{(\rm II, IIn)}_\nu(E_\nu, D_*)/dE_\nu$ being the total emitted neutrino spectra from individual Type II and Type IIn SN with a given $D_*$. We adopt $R_{\rm csm}=4\times 10^{14}\,\mathrm{cm}$ for Type II SNe and $R_{\rm csm}=10^{16}\,\mathrm{cm}$ for Type IIn SNe, and consider their normalized $D_*$ distributions $f_{\rm II}$ and $f_{\rm IIn}$ (Fig.~1), weighted by the relative fractions $\eta_{\rm II}\simeq61.3\%$ and $\eta_{\rm IIn}\simeq5.7\%$ of all CCSNe inferred from recent SN observations~\cite{Ma:2025css}. We further fix $\epsilon_p=0.1$ and consider $\epsilon_B=10^{-4}$ and 0.1.

\begin{figure}[htbp]
\centering
\includegraphics[width=\columnwidth]{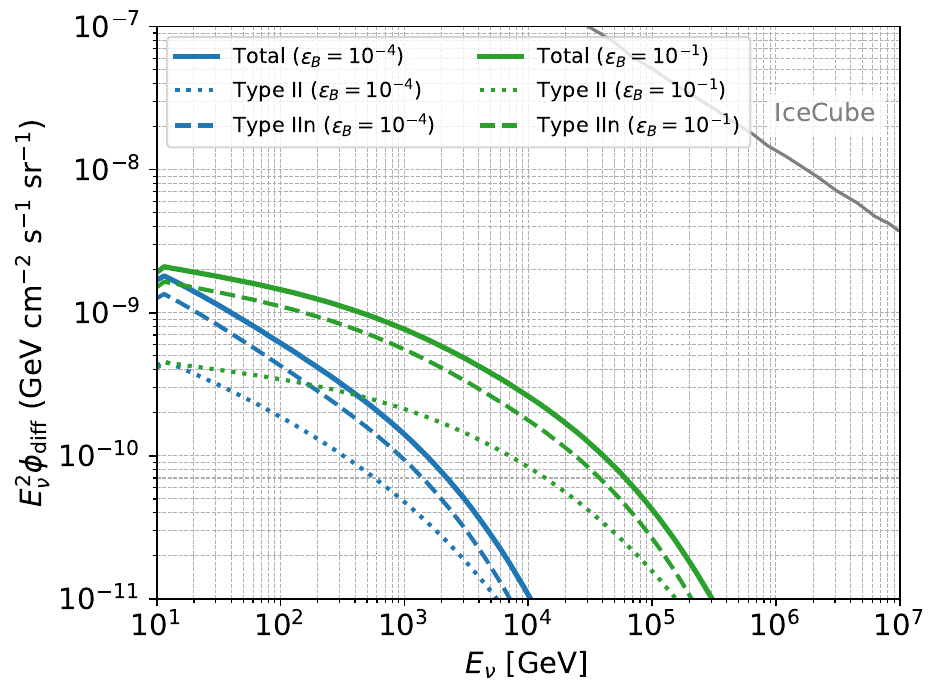}
\caption{Predicted diffuse HE neutrino flux from Type II and Type IIn SNe in the SN ejecta-CSM interaction model. Contributions from Type~II and Type~IIn SNe, as well as their sum, are shown for $\epsilon_B = 10^{-4}$ (blue) and 0.1 (green). The detected IceCube 10-year diffuse neutrino flux using starting track events \cite{IceCube:2024fxo} is shown for comparison (grey).}
\label{fig:diffuse_flux}
\end{figure}

Figure~\ref{fig:diffuse_flux} shows the resulting diffuse neutrino flux, including the separate contributions from Type~II and Type~IIn SNe, together with their sum. As discussed above, the $D_*$ distribution of regular Type II SNe, when weighted by their relative occurrence rates, exceeds that of Type IIn SNe, even at the high-$D_*$ end. However, typical Type II SNe are surrounded by more compact CSM. As a result, the diffuse neutrino flux is dominated by Type IIn SNe, in agreement with earlier studies \cite{Ashida:2024nck}. Nevertheless, regular Type II SNe provide a non-negligible contribution, particularly at the highest energies.

Even under optimistic assumptions ($\epsilon_B=0.1$), the predicted diffuse flux arising from the SN ejecta--CSM interactions remains 2--3 orders of magnitude below the observed flux by IceCube (the grey line) \cite{IceCube:2024fxo}, using the 10.3-year Enhanced Starting Track Event Selection (ESTES) dataset. The predicted flux is also consistent with the limits derived from IceCube searches for HE neutrinos from samples of known CCSNe \cite{IceCube:2023esf}. This small diffuse contribution implies that HE neutrinos from nearby individual SNe offer more promising detection prospects.

\section{Constraining CSM and Particle Acceleration Parameters}
\label{sec:constrain}

In the previous section, we assessed the discovery significance of HE neutrinos from a single nearby SN by adopting fixed values for the parameter set $\boldsymbol{\lambda} = (D_*, R_{\rm csm}, \epsilon_B, \epsilon_p)$. In practice, however, these parameters are all highly uncertain and may vary substantially across different SN progenitors. In this section, we investigate the potential of using HE neutrino signals to constrain these parameters, incorporating their uncertainties within a consistent statistical framework.

Since the expected background rate is low, we neglect its uncertainty and define the likelihood function as
\begin{align}
\mathcal{L}'(\boldsymbol{\lambda}) = \prod_{ij} P(N_{ij}, s_{ij}(\boldsymbol{\lambda}) + b_{ij}),
\label{eq:likeli2}
\end{align}
where all quantities are defined as in Eq.~\eqref{eq:likeli}. To estimate the median expected sensitivity, we employ the Asimov dataset, for which the observed counts are fixed to $N_{ij}=s_{ij}(\boldsymbol{\lambda}_0)+b_{ij}$,
with $\boldsymbol{\lambda}_0$ denoting a benchmark parameter set. For this dataset, the global maximum of the likelihood naturally occurs at 
$\boldsymbol{\lambda}=\boldsymbol{\lambda}_0$. 

For illustrative purpose, we project constraints in both one- and two-dimensional parameter spaces using the above defined likelihood function. For one-dimensional constraints, a single parameter of interest is varied while the remaining three parameters are allowed to float freely within physically motivated ranges. For example, when constraining the parameter $D_*$, the TS is defined as
\begin{align}
\mathrm{TS}'(D_*) =
2 \left[
\ln \mathcal{L}'\!\left(D_*,\, \widehat{R}_{\rm csm},\, \widehat{\epsilon}_B,\, \widehat{\epsilon}_p\right)
-
\ln \mathcal{L}'(\boldsymbol{\lambda}_0)
\right],
\label{eq:TS1D}
\end{align}
where $\widehat{R}_{\rm csm}$, $\widehat{\epsilon}_B$, and $\widehat{\epsilon}_p$ denote the values that maximize the likelihood for each fixed $D_*$. Similarly, for two-dimensional constraints, the likelihood is evaluated for a pair of parameters of interest, with the remaining two parameters profiled over. For instance, when constraining $(D_*, R_{\rm csm})$, the TS is
\begin{align}
\mathrm{TS}'(D_*, R_{\rm csm}) =
2 \left[
\ln \mathcal{L}'\!\left(D_*, R_{\rm csm}, \widehat{\epsilon}_B, \widehat{\epsilon}_p\right)
-\ln \mathcal{L}'(\boldsymbol{\lambda}_0)\right],
\end{align}
where $\widehat{\epsilon}_B$ and $\widehat{\epsilon}_p$ maximize the likelihood at each fixed $(D_*, R_{\rm csm})$. The resulting TS contours quantify the joint constraints on the parameters and reveal correlations among them.

\begin{figure*}[htbp]
\centering
\includegraphics[width=\columnwidth]{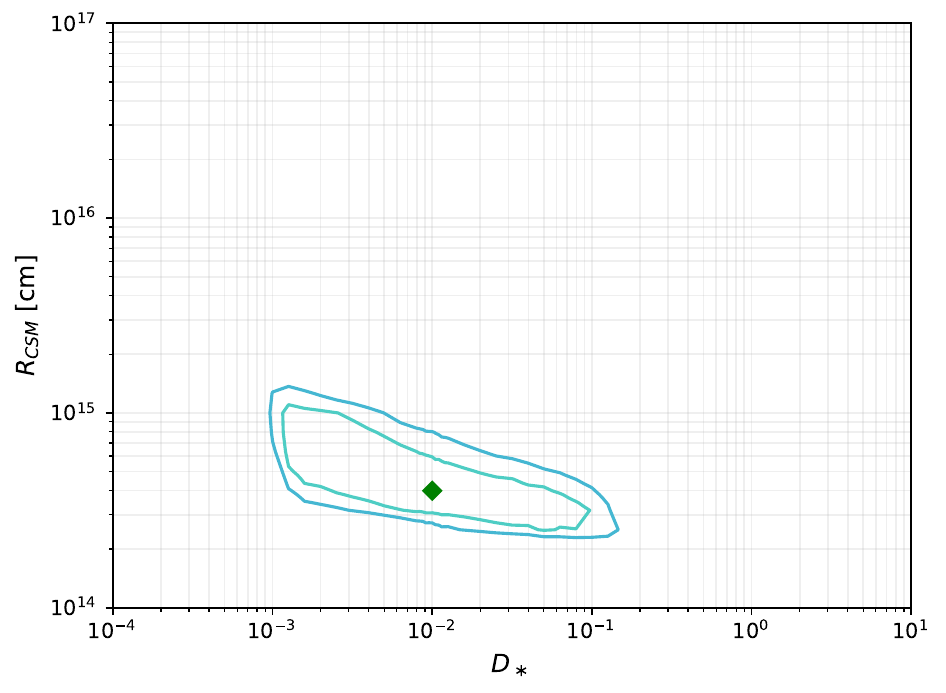}
\includegraphics[width=\columnwidth]{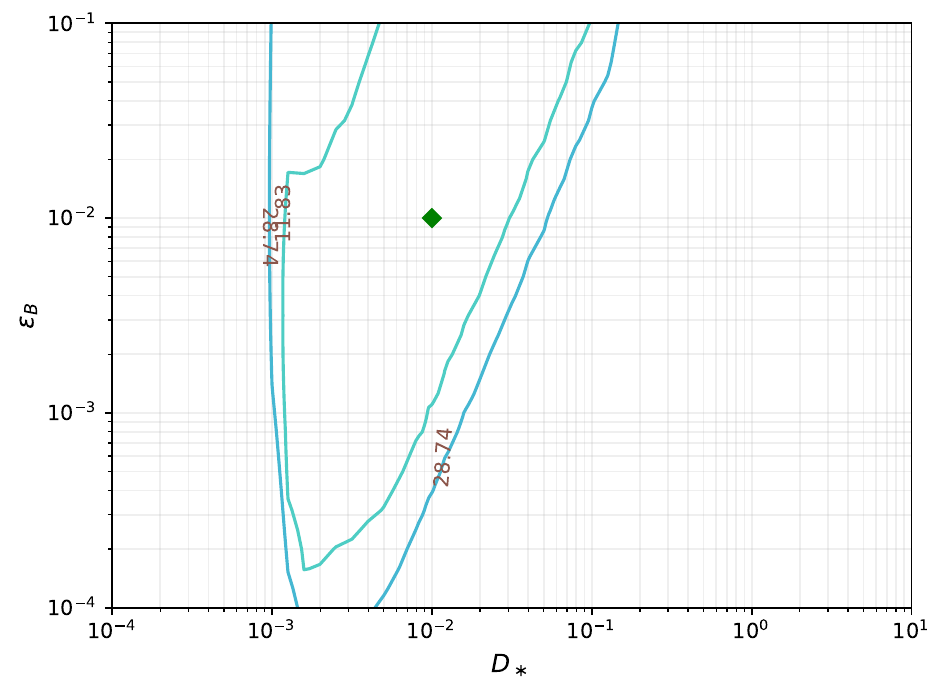}  \\
\includegraphics[width=\columnwidth]{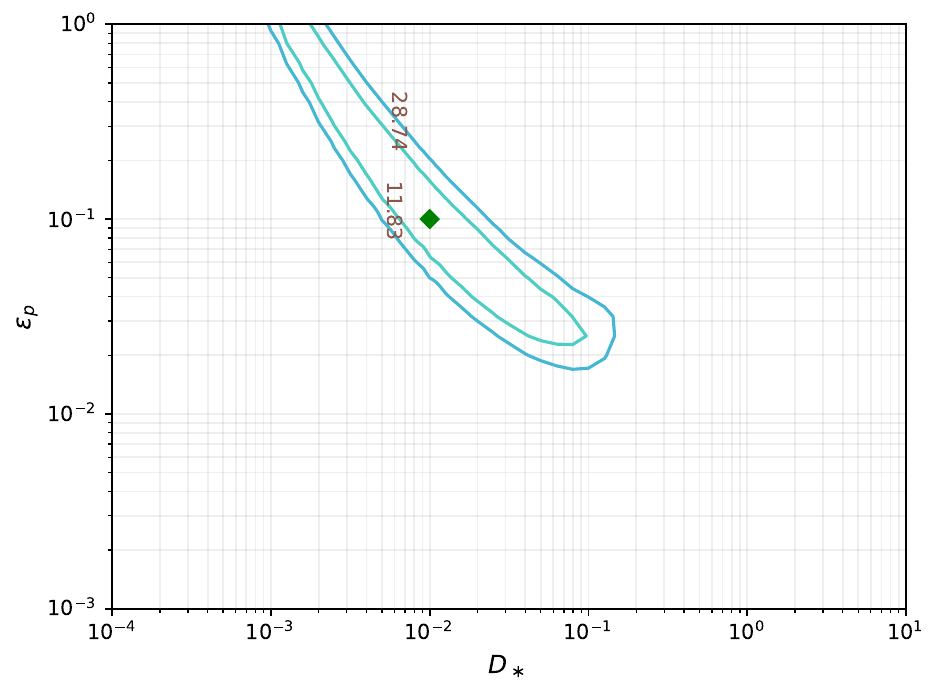}
\includegraphics[width=\columnwidth]{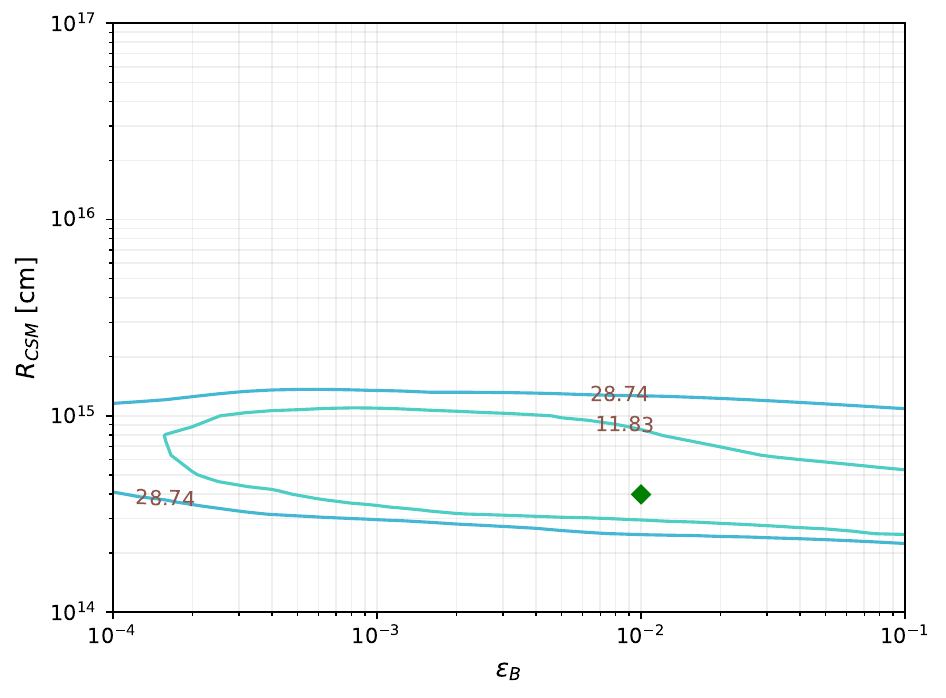} \\
\includegraphics[width=\columnwidth]{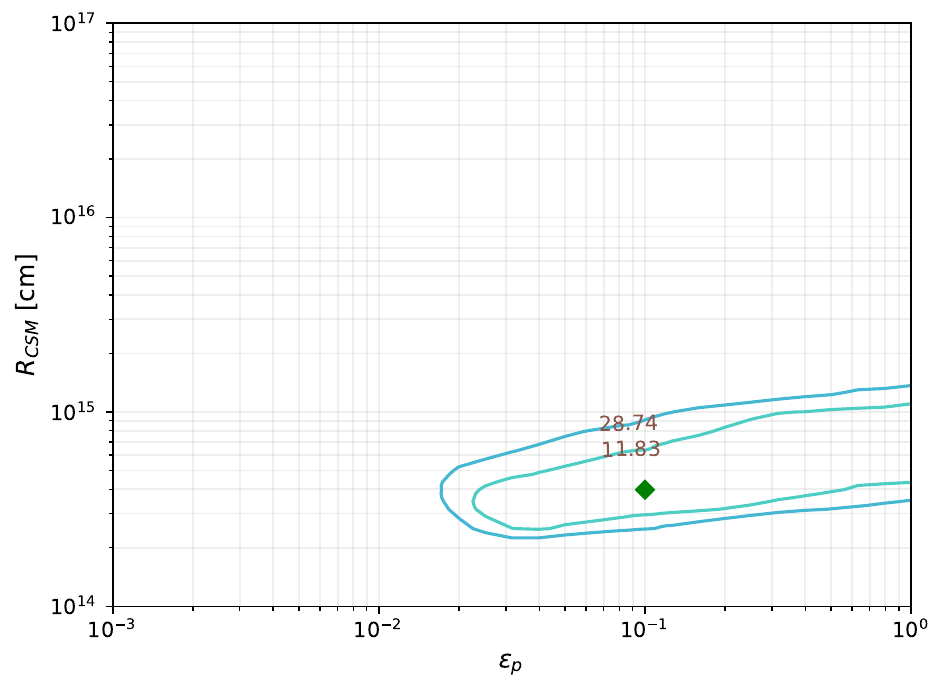}
\includegraphics[width=\columnwidth]{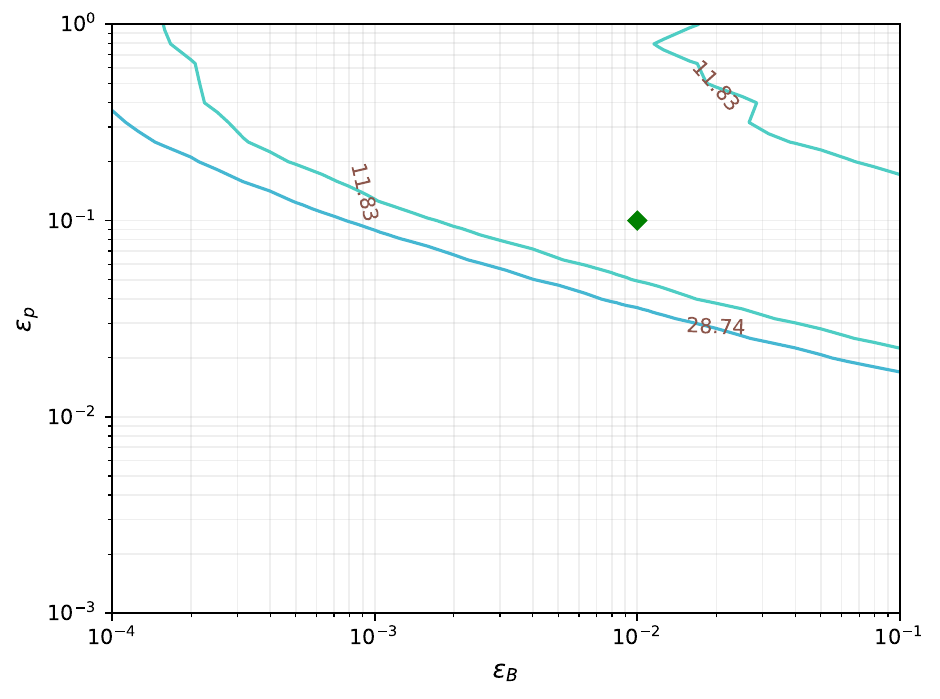}
\caption{Two-dimensional profile likelihood constraints on the parameters for a Galactic Type II SN at 10~kpc. Shown are $3\sigma$ (${\rm TS}=11.83$) and $5\sigma$ (${\rm TS}=28.74$) contours of in the planes of $(D_*,R_{\rm csm})$ (upper left), $(D_*,\epsilon_B)$ (upper right), $(D_*,\epsilon_p)$ (middle left), $(\epsilon_B,R_{\rm csm})$ (middle right), $(\epsilon_p,R_{\rm csm})$ (lower left), and $(\epsilon_B,\epsilon_p)$ (lower right). In each panel, the two parameters shown are varied while the other two parameters are profiled over. For the benchmark model, we consider a typical Type II SN located at 10 kpc with $D_*=0.01$, $R_{\rm csm}=4\times 10^{14}\,\mathrm{cm}$, $\epsilon_B=0.01$, and $\epsilon_p=0.1$ indicated by filled diamond symbols in the 2D parameter spaces.}
\label{fig:2Dconstraint}
\end{figure*}

For our analysis, the parameters are varied over typical ranges inferred from observations of Type II SNe:
\begin{align}
& D_*\in[10^{-4}, 10],\quad R_{\rm csm}\in[10^{14},10^{17}]~\mathrm{cm}, \nonumber \\
& \epsilon_B\in[10^{-4},0.1],\quad \epsilon_p\in[10^{-4},1]. \nonumber 
\end{align}
Following Wilks' theorem, $\mathrm{TS}$ values of 9 and 25 correspond to 3$\sigma$ and 5$\sigma$ CLs, respectively, for one-dimensional constraints, while the corresponding values for two-dimensional constraints are ${\rm TS} \approx 11.83$ and 28.74.

\begin{figure*}[htbp]
\centering
\includegraphics[width=\columnwidth]{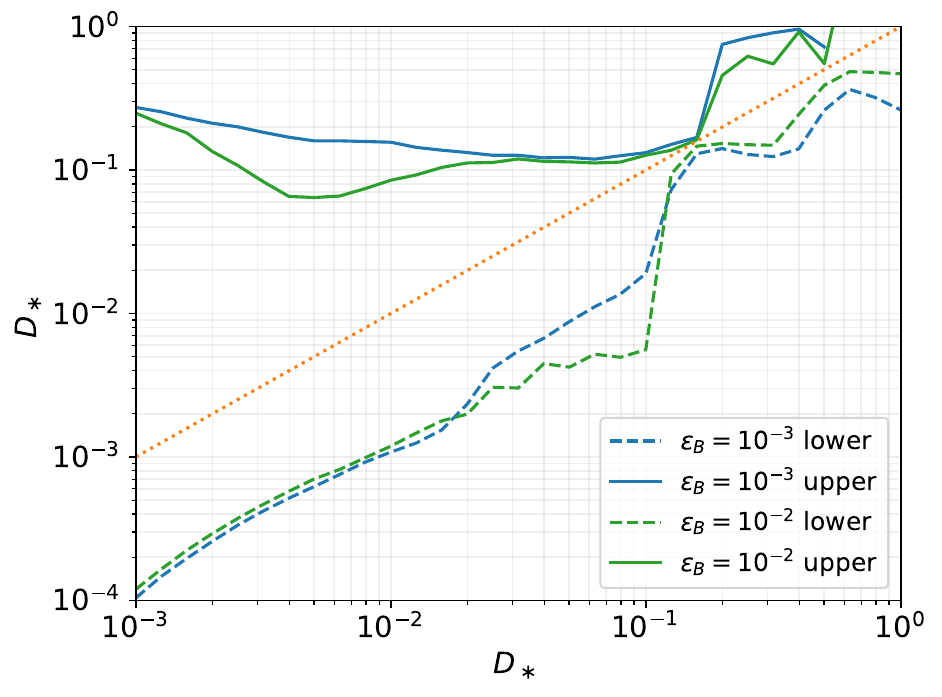}
\includegraphics[width=\columnwidth]{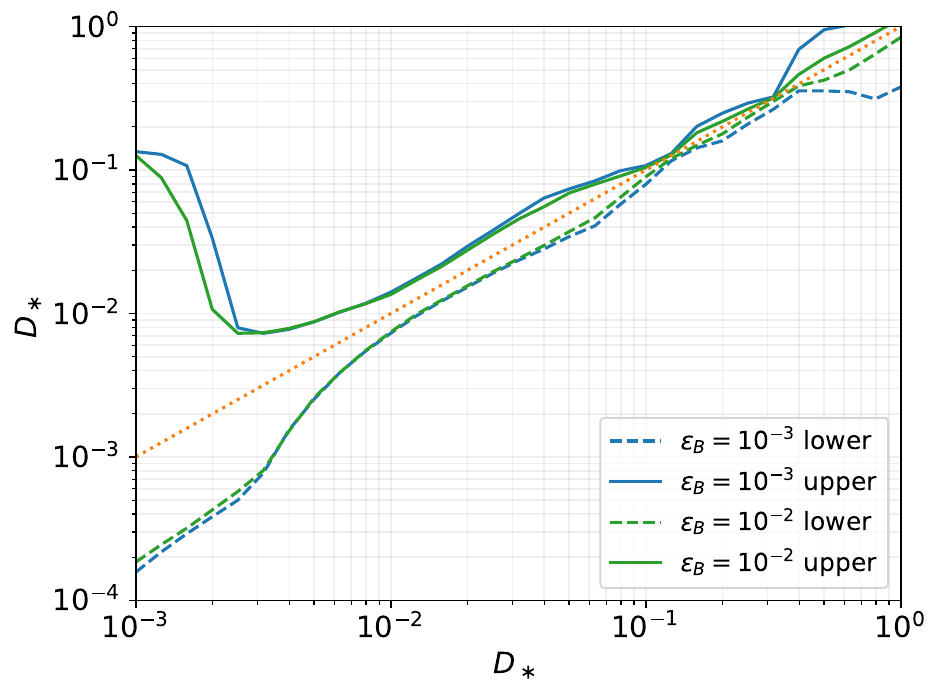}  \\
\includegraphics[width=\columnwidth]{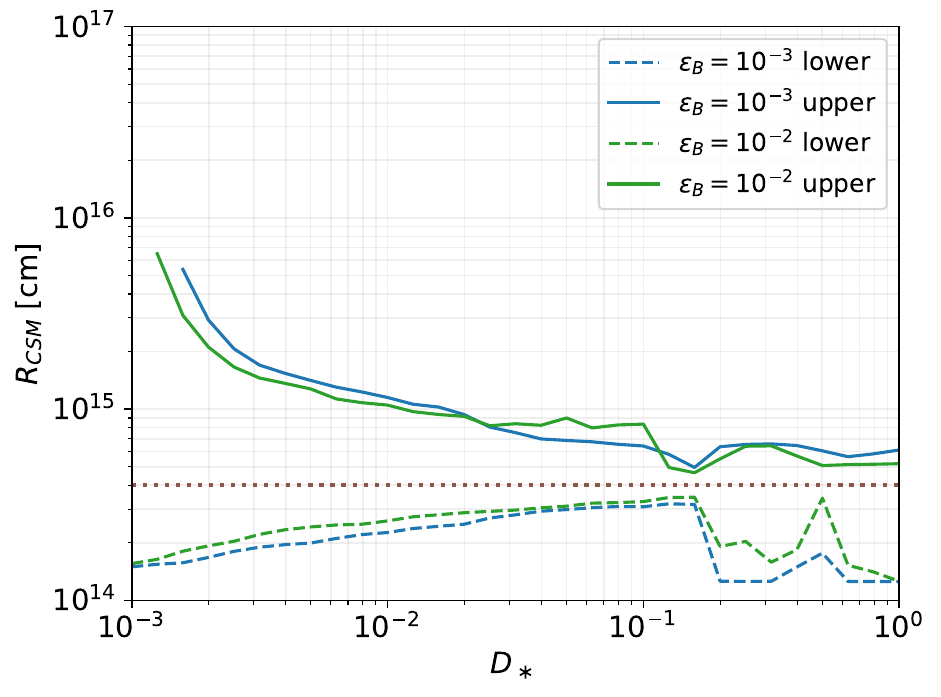}
\includegraphics[width=\columnwidth]{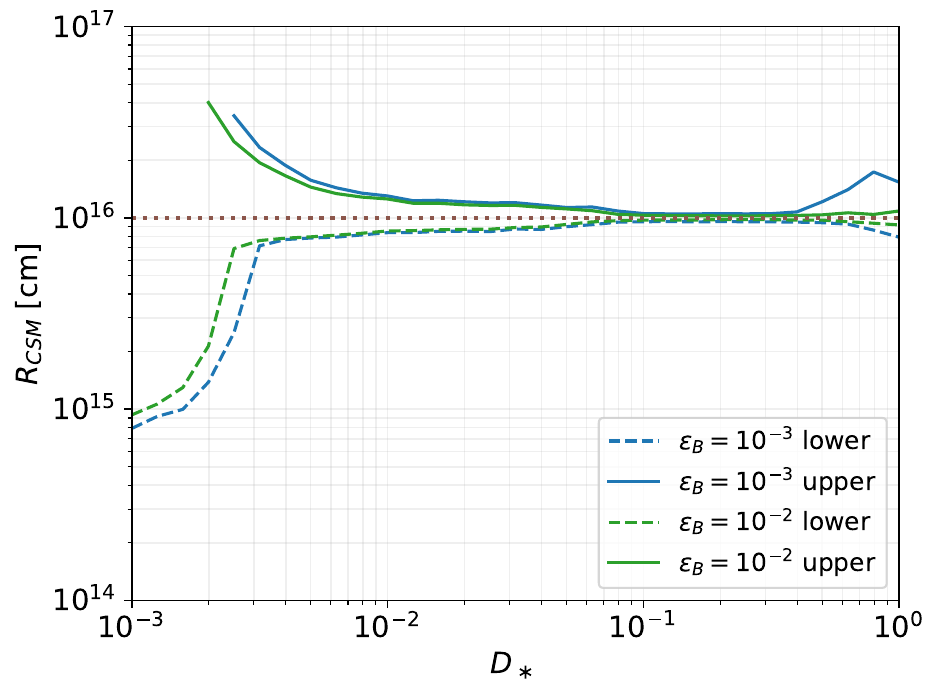}
\caption{One-dimensional constraints on the parameters $D_*$ (upper) and $R_{\rm csm}$ (lower) for a Galactic Type II (left) and Type IIn (right) SN at distances of 10~kpc. The $x$-axis shows the $D_*$ values of the benchmark model, with $R_{\rm csm}=4\times 10^{14}$~cm (Type II) and $10^{16}$~cm (Type IIn), assuming $\epsilon_B=10^{-3}$, $10^{-2}$ and $\epsilon_p=0.1$. The $y$-axis denotes the projected 3$\sigma$ lower and upper limits on $D_*$ and $R_{\rm csm}$ using the HE neutrino signals.}
\label{fig:1Dconstraint_DR}
\end{figure*}

The two-dimensional constraints for a Type II SN at a distance of 10 kpc are shown in Fig.~\ref{fig:2Dconstraint}. For illustration, we adopt the benchmark values $D_*=0.01$,
$R_{\rm csm}=4\times 10^{14}$~cm, $\epsilon_B=0.01$, and $\epsilon_p=0.1$. The most prominent correlation appears in the $(D_*,\epsilon_p)$ plane, where the confidence contours exhibit an elongation along a diagonal direction, indicating that the combination $D_*/\epsilon_p$ is relatively well constrained over a certain parameter region. The degeneracy between $D_*$ and $\epsilon_p$ is physically expected, as both parameters influence the overall signal strength. However, variations in $D_*$ introduce distinct spectral and temporal features that cannot be fully reproduced by rescaling $\epsilon_p$ alone, thereby partially breaking the degeneracy once full energy- and time-dependent information is incorporated in the likelihood analysis. Comparably weaker correlations are present in the $(D_*,R_{\rm csm})$ and $(R_{\rm csm},\epsilon_p)$ planes for similar reasons. The parameter $\epsilon_B$ mainly affects the HE cutoff of the spectrum. This effect can be partially mimicked by correlated changes in other parameters: as variations in $D_*$ can modify the spectral shape while $\epsilon_p$ rescales the overall neutrino flux. Consequently, mild residual degeneracies arise between $\epsilon_B$ and $D_*$ or $\epsilon_p$. There is nearly no degeneracy between $R_{\rm csm}$ and $\epsilon_B$.

As shown in the Fig.~\ref{fig:2Dconstraint}, for Type~II SNe with $D_*=0.01$, the parameter $R_{\rm csm}$ can be constrained to within a factor of $\sim 2$--3. By comparison, $D_*$ and $\epsilon_p$ can only be constrained to within a factor of $\sim 10$, while $\epsilon_B$ remains essentially unconstrained. These differences reflect the distinct roles of the model parameters in shaping the neutrino signal. The parameters $D_*$, $R_{\rm csm}$, and $\epsilon_p$ all control the overall signal strength (see Sec.~\ref{sec:basic}); however, $R_{\rm csm}$ additionally governs the temporal evolution of the emission, in particular its duration and termination, which cannot be easily reproduced by variations of other parameters. With the inclusion of time-dependent information in the likelihood analysis, $R_{\rm csm}$ can therefore be relatively well constrained. In contrast, although $D_*$ also affects the neutrino spectrum, its impact on the flux normalization and spectral cutoff can be partially mimicked by correlated variations in $\epsilon_p$ and $\epsilon_B$. This degeneracy (see the middle left panel of Fig.~\ref{fig:2Dconstraint}) limits the constraining power on both $D_*$ and $\epsilon_p$. The parameter $\epsilon_B$ primarily influences the HE cutoff and has a comparatively modest effect on the overall signal, resulting in weaker constraints.

\begin{figure*}[htbp]
\centering
\includegraphics[width=\columnwidth]{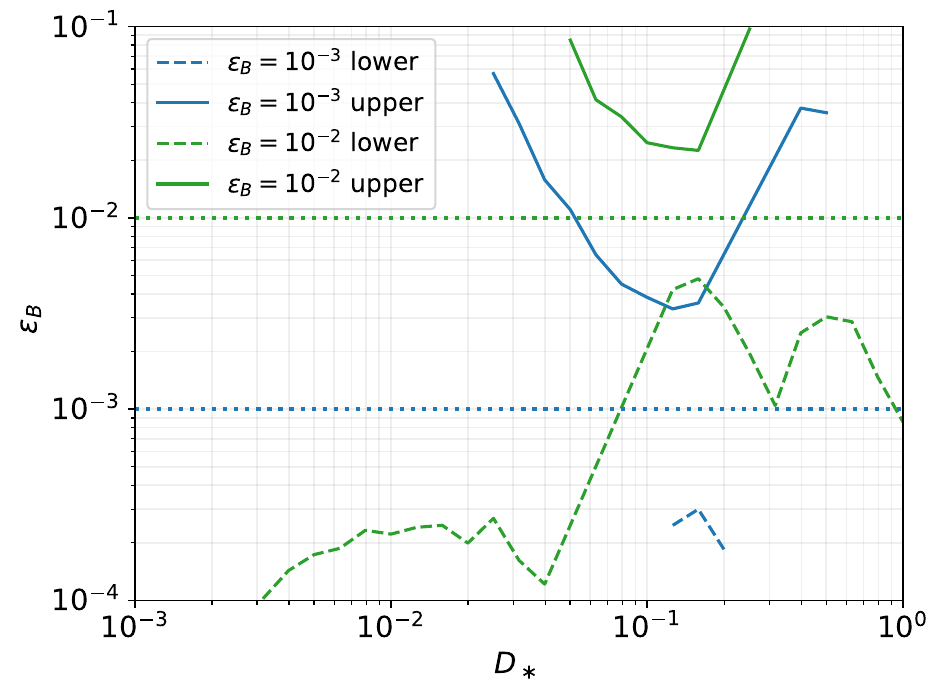}
\includegraphics[width=\columnwidth]{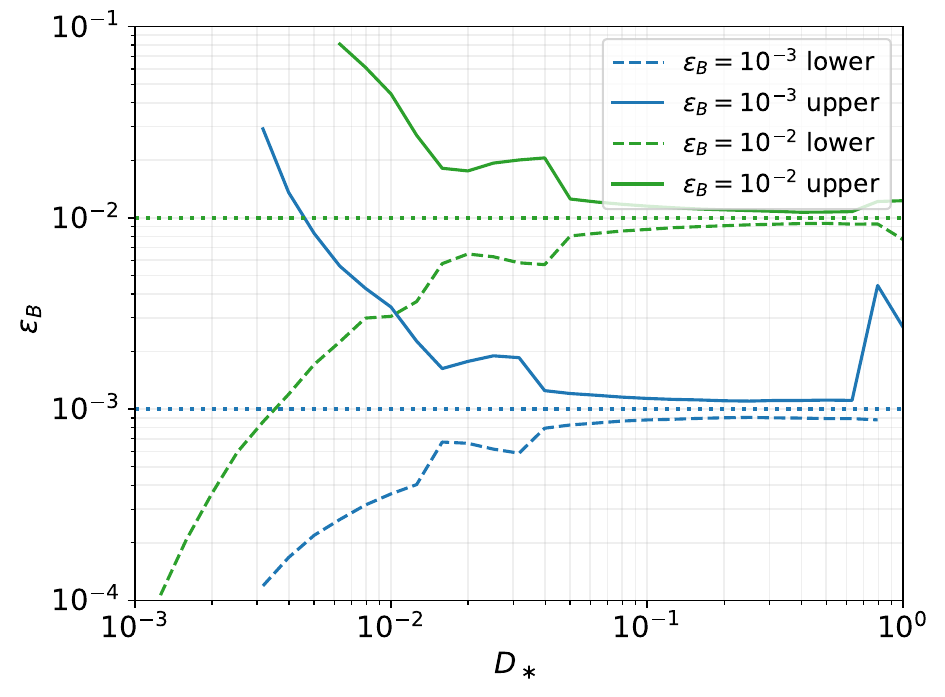}  \\
\includegraphics[width=\columnwidth]{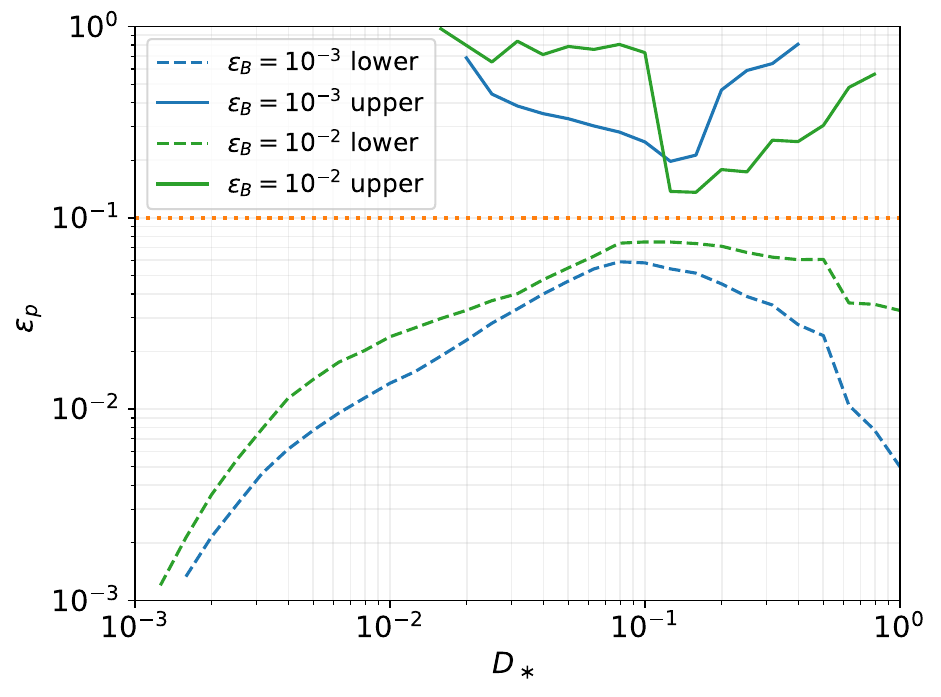}
\includegraphics[width=\columnwidth]{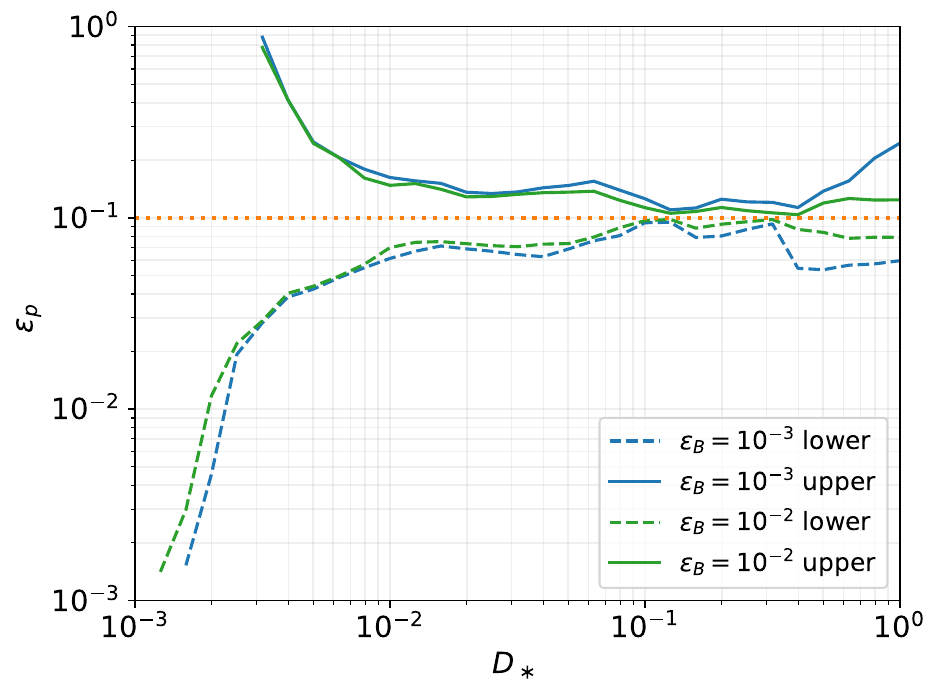}
\caption{Same as Fig.~\ref{fig:1Dconstraint_DR}, but showing the constraints on $\epsilon_B$ (upper) and $\epsilon_p$ (lower) for Type II (left) and Type IIn (right) SNe.}
\label{fig:1Dconstraint_Bp}
\end{figure*}

To account for the diversity of SN populations, we present the projected one-dimensional constraints (3$\sigma$ CL) on each relevant parameter for both Type~II ($R_{\rm csm}=4\times10^{14}\,\mathrm{cm}$) and Type~IIn ($R_{\rm csm}=10^{16}\,\mathrm{cm}$) SNe at distances of 10 kpc in Figs.~\ref{fig:1Dconstraint_DR} (for $D_*$ and $R_{\rm csm}$) and \ref{fig:1Dconstraint_Bp} (for $\epsilon_B$ and $\epsilon_p$), assuming different benchmark values of $D_*$. Two representative values, $\epsilon_B=10^{-3}$ and $10^{-2}$, are considered. For each parameter, the corresponding benchmark value is indicated by a diamond symbol. The comparison between the projected upper and lower bounds and the benchmark values illustrates how well each parameter can be constrained for a given benchmark model.

As $D_*$ increases, the expected number of signal events grows, leading to progressively tighter constraints on all relevant parameters (Figs.~\ref{fig:1Dconstraint_DR} and \ref{fig:1Dconstraint_Bp}). At the optimal benchmark value $D_* \sim 0.1$ for Type~II SNe, $D_*$, $R_{\rm csm}$, and $\epsilon_p$ can all be constrained to within a factor of $\sim 2$, while $\epsilon_B$ can be determined to within a factor of $\sim 5$--10. We note that high event statistics significantly reduce the impact of parameter degeneracies, allowing $R_{\rm csm}$, $D_*$, and $\epsilon_p$ to be constrained with similar precision.

As shown in the right panels of Figs.~\ref{fig:1Dconstraint_DR} and \ref{fig:1Dconstraint_Bp}, the constraining power is further improved for SNe surrounded by extended CSM (Type~IIn) at the same distance. For benchmark values of $D_*$ from $\sim 3\times10^{-3}$ to $\sim 0.5$, $D_*$, $R_{\rm csm}$, and $\epsilon_p$ can be measured with a precision of $\sim 20$--30\%. Furthermore, for $D_*$ between $\sim 0.03$ and $\sim 0.7$, $\epsilon_B$ can also be constrained to a comparable level. These results are largely insensitive to the assumed benchmark values of $\epsilon_B$ ($10^{-3}$ and $10^{-2}$).

Overall, despite the large uncertainties and broad ranges of the relevant parameters, HE neutrino observations of a nearby SN with sufficient statistics and distinctive spectral and temporal features can constrain the CSM profile ($D_*$ and $R_{\rm csm}$) and the proton acceleration efficiency $\epsilon_p$, while providing comparatively weaker but still meaningful constraints on $\epsilon_B$. The modest parameter degeneracies identified in this analysis do not significantly compromise these conclusions, especially when the event statistics are sufficiently large.

\section{Discussion and Conclusion}
\label{sec:summary}

In this work, we study HE neutrino production from interactions between the SN ejecta and the surrounding CSM, following the formalism of Refs.~\cite{Murase:2017pfe,Pitik:2023vcg}. We focus on regular Type~II and Type~IIn SNe, adopting observationally inferred distributions of the CSM density normalization $D_*$ from large SN samples, as well as characteristic CSM extents $R_{\rm csm}$. Within this framework, we explore the impact of key physical parameters---including $D_*$, $R_{\rm csm}$, the proton acceleration efficiency $\epsilon_p$, and the magnetic energy fraction $\epsilon_B$---on the resulting neutrino flux. We also compare calculations with and without explicitly solving the time-dependent transport of accelerated protons, finding that proton transport introduces modest but non-negligible modifications to the predicted HE neutrino signal.

We evaluate the detectability of these neutrinos using a binned likelihood approach in both time and energy, focusing on upgoing track events at IceCube. The detection prospects depend sensitively on the CSM profile and shock microphysics. For demonstration, we fix $\epsilon_p=0.1$, adopt $R_{\rm csm}=4\times 10^{14}$~cm and $10^{16}$~cm for Type II and Type IIn SNe, respectively, and consider two representative values of $\epsilon_B=10^{-4}$ and 0.1. We find an optimal density range $D_* \sim 0.03$--0.3 (0.1--0.5) for typical Type II (IIn) SNe, for which the detection horizon extends to $\sim 0.05$--0.2~Mpc (0.15--0.6 Mpc), depending on $\epsilon_B$ ($10^{-4}$--0.1). For Galactic SNe at distances of $\sim 10$~kpc, a broad range of CSM densities ($D_* \sim 10^{-3}$--$10$) can yield detectable signals at $\ge 3\sigma$. These results indicate that nearby regular Type~II SNe with moderately dense, confined CSM are promising targets for HE neutrino observations, while Type~IIn SNe represent the most luminous but rarer sources.

By integrating the neutrino emission over the cosmic CCSN rate, we also evaluate the diffuse neutrino flux contributed by the entire populations of Type II and Type IIn SNe. We find that ordinary Type II SNe make a non-negligible contribution to the diffuse flux, owing to their substantially higher event rate and the existence of a tail of systems exhibiting enhanced, short-lived mass loss. Nevertheless, even under optimistic assumptions, the predicted diffuse flux remains well below the diffuse astrophysical neutrino intensity measured by IceCube. This indicates that targeted searches for neutrinos from nearby SNe are considerably more promising \cite{Kheirandish:2022eox}.

A key result of this work is that neutrino observations from a future Galactic Type~II or Type~IIn SN, with detailed spectral and temporal information, can place meaningful constraints on the CSM density profile and shock-acceleration parameters, despite their broad physically allowed ranges. Using profile-likelihood analyses for a Type~II SN with $D_* = 0.01$ (and Type~IIn SN with $D_* \sim 0.01$--1) at 10~kpc, we find that $R_{\rm csm}$ can be constrained to within a factor of $\sim 2$ (20\%). By comparison, $D_*$ and $\epsilon_p$ can be measured within a factor of $\sim 10$ (20\%), while $\epsilon_B$ is weakly constrained for Type~II SNe and to $\sim 20\%$ for Type~IIn SNe. The constraining power increases with $D_*$, as shown in Figs.~\ref{fig:1Dconstraint_DR} and \ref{fig:1Dconstraint_Bp}. Parameter degeneracies can reduce precision, but this effect is largely mitigated at higher event statistics corresponding to larger $D_*$ or extended CSM, preserving the sensitivity of the neutrino signal. Although not explored in detail in this work, next-generation neutrino observatories such as IceCube-Gen2, KM3NeT, and TRIDENT will further extend the detection horizon and enhance parameter constraints.

We fix $M_{\rm ej}$ and $\mathcal{E}_{\rm ej}$ in our study, but variations in these quantities can be effectively absorbed into $D_*$ [Eq.~\eqref{eq:Rs}]. We also adopt a canonical wind-like CSM profile with density slope $w=2$ and a fixed proton acceleration index $p=2$. While these assumptions may not apply to all progenitors, high-statistics neutrino observations with detailed temporal and spectral information are expected to remain sensitive to a broader parameter space. Overall, within the framework of a simple ejecta--CSM interaction model, we demonstrate that HE neutrino observations of a Galactic SN, with sufficient statistics and detailed energy--time information, can provide a meaningful and complementary probe of the CSM density profile and shock acceleration, beyond what is accessible with traditional electromagnetic observations.

\begin{acknowledgments}
This work was supported in part by the National SKA Program of China (No.~2025SKA0110104), Guangdong Basic and Applied Basic Research Foundation (No.~2025A1515011082), and the ``CUG Scholar" Scientific Research Funds at China University of Geosciences (Wuhan) [No.~2021108].
\end{acknowledgments}

% 参考文献
%\bibliographystyle{apsrev4-2}
%\bibliography{references} % references.bib 文件应该包含你的所有引用


\end{document}